\documentclass[preprint,prd,aps,showpacs,showkeys,nofootinbib]{revtex4}
\usepackage{graphicx}
\usepackage{dcolumn}
\usepackage{bm}
\usepackage{color}
\usepackage[dvipsnames]{xcolor}
\usepackage{amssymb}

\definecolor{light-gray}{gray}{0.78}
\definecolor{mid-gray}{gray}{0.55}
\definecolor{dark-gray}{gray}{0.32}

\begin{document}
\title{Higgs Bosons at 95 and 125 GeV in the $U(1)_X$VLFM}
\author{Rong-Zhi Sun$^{1,2,3}$, Shu-Min Zhao$^{1,2,3}$\footnote{zhaosm@hbu.edu.cn}, Meng-Zi Cao$^{1,2,3}$, Song Gao$^{1,2,3}$, Xing-Xing Dong$^{1,2,3,4}$\footnote{dongxx@hbu.edu.cn}}
\affiliation{$^1$ Department of Physics, Hebei University, Baoding 071002, China}
\affiliation{$^2$ Hebei Key Laboratory of High-precision Computation and Application of Quantum Field Theory, Baoding, 071002, China}
\affiliation{$^3$ Hebei Research Center of the Basic Discipline for Computational Physics, Baoding, 071002, China}
\affiliation{$^4$ Departamento de Fisica and CFTP, Instituto Superior T$\acute{e}$cnico, Universidade de Lisboa,
Av.Rovisco Pais 1,1049-001 Lisboa, Portugal}
\date{\today}

\begin{abstract}
We present a systematic analysis of the Higgs signal strengths at 125 GeV and 95 GeV in a non-supersymmetric $U(1)_X$ model with vector-like fermions ($U(1)_X$VLFM). This model extends the SM by introducing an additional $U(1)_X$ gauge symmetry, three right-handed neutrinos, two singlet Higgs fields ($\phi$ and $S$), and one generation of vector-like quarks and leptons. The scalar fields mix with each other in the neutral CP-even sector, leading to two Higgs-like states around 95 GeV and 125 GeV. A $\chi^2$ analysis is performed by combining the Higgs signal strength measurements at 125 GeV from ATLAS and CMS, including the $\gamma\gamma$, $WW^*$, $ZZ^*$, $b\bar{b}$, and $\tau\bar{\tau}$ channels, with the 95 GeV excesses observed in the diphoton and $b\bar{b}$ final states reported by CMS and LEP. Our results indicate that the $U(1)_X$VLFM can successfully reproduce the observed signal strengths of the 125 GeV Higgs while simultaneously explaining the 95 GeV excess. The parameters $g_X$, $g_{YX}$, $v_S$, $v_P$, and the new Yukawa couplings play a crucial role in achieving this consistency.
\end{abstract}

\maketitle
\section{Introduction}
Since the discovery of a new scalar particle with a mass of about 125 GeV by the ATLAS and CMS collaborations at the Large Hadron Collider (LHC) in 2012~\cite{ATLAS:2012yve,CMS:2012qbp,ATLAS:2016neq}, its observed properties have been consistent within current theoretical and experimental uncertainties, with those expected for the Standard Model (SM) Higgs boson. Nevertheless, such a state can also be accommodated in a wide range of theories beyond the Standard Model (BSM). Although no direct evidence for BSM physics has yet emerged at the LHC, the precision of Higgs coupling measurements and the constraints from searches for new resonances still allow substantial parameter space for new physics (NP) interpretations. Many BSM scenarios predict an extended Higgs sector, making it an essential task for the LHC Run 3 and future experiments to determine whether the discovered scalar boson is part of a richer Higgs structure. Notably, these extended frameworks may feature not only heavier Higgs resonances but also lighter scalar states. The search for such light additional Higgs bosons is therefore of great importance for uncovering the underlying mechanism of electroweak symmetry breaking and possible signs of NP.

The current mass measurement for the 125 GeV Higgs boson is $125.20 \pm 0.11$ GeV~\cite{PDG}. To investigate its properties in detail and to assess whether it is the only fundamental scalar, the LHC has performed precision measurements across multiple decay channels. In particular, the CMS and ATLAS collaborations have observed the Higgs in various bosonic and fermionic modes~\cite{PDG,gamma1,gamma2,ATLAS:2016neq}, establishing its spin-parity quantum numbers and measuring its production cross sections. In the SM, the Higgs couples directly to $W$ and $Z$ bosons and indirectly to photons via loop effects, making the $\gamma\gamma$, $WW^*$, and $ZZ^*$ channels particularly sensitive for experimental studies. The measured signal strengths in these channels are~\cite{gamma1,gamma2,ATLAS:2016neq,gamma4,zz}
\begin{equation}
\mu_{\gamma\gamma}^{exp}(125) = 1.10 \pm 0.06, \quad
\mu_{WW^*}^{exp}(125) = 1.00 \pm 0.08, \quad
\mu_{ZZ^*}^{exp}(125) = 1.02 \pm 0.08.
\end{equation}
Owing to Yukawa interactions, the Higgs couples most strongly to third-generation fermions, resulting in dominant decays into $b\bar b$ and $\tau\bar\tau$. The corresponding measured signal strengths are~\cite{gamma2,ATLAS:2016neq,gamma4,bb1,bb2,tautau}
\begin{equation}
\mu_{b\bar b}^{exp}(125) = 0.99 \pm 0.12, \quad
\mu_{\tau\bar\tau}^{exp}(125) = 0.91 \pm 0.09.
\end{equation}

Besides the established Higgs boson at 125 GeV, several experimental studies have pointed to a possible lighter scalar resonance near 95 GeV, renewing interest in extended Higgs sectors. Searches for additional low-mass Higgs states have been carried out at LEP \cite{OPAL:2002ifx,LEPWorkingGroupforHiggsbosonsearches:2003ing,ALEPH:2006tnd}, Tevatron \cite{CDF:2012wzv}, and the LHC \cite{CMS:2015ocq,CMS:2018cyk,CMS:2018rmh,ATLAS:2018xad,CMS:2022goy,ATLAS:2022abz,CMS:2023yay}. For new Higgs boson searches, the experimental signal strength $\mu$ is defined relative to a hypothetical SM Higgs boson of the same mass, even if no SM Higgs exists at that mass. Specifically, the experimental signal strengths at $m_\phi = 95.4~\rm{GeV}$ are
\begin{eqnarray}
&&
\mu_{\gamma\gamma}^{\rm exp} = \frac{\sigma^{\rm exp}(gg\to\phi\to\gamma\gamma)}{\sigma^{\rm SM}(gg\to H_{\rm SM}\to\gamma\gamma)},\qquad\nonumber\\&&
\mu_{b\bar b}^{\rm exp} = \frac{\sigma^{\rm exp}(e^+e^-\to Z\phi\to Zb\bar b)}{\sigma^{\rm SM}(e^+e^-\to Z H_{\rm SM}\to Zb\bar b)},
\end{eqnarray}
where $\phi$ denotes a hypothetical BSM scalar, and $\sigma^{\rm SM}$ is the cross section for a SM-like Higgs of the same mass~\cite{Biekotter:2023oen,Chen:2023bqr}.

The CMS Collaboration analyses in the diphoton channel at $\sqrt{s}=8$ TeV and 13 TeV, with datasets of 19.7 fb$^{-1}$ and 35.9 fb$^{-1}$, revealed a local excess around 95.3 GeV with a significance of about 2.8$\sigma$ \cite{CMS:2015ocq,CMS:2018cyk}. Using the full Run 2 dataset and improved event classification, a more recent CMS study reports a 2.9$\sigma$ local excess at 95.4 GeV \cite{CMS:2023yay}, broadly compatible with ATLAS results based on 80 fb$^{-1}$ of data \cite{ATLAS:2018xad}. The combined ATLAS + CMS signal strength can be expressed as
\begin{equation}
\mu_{\gamma\gamma}^{exp}(95) = 0.24^{+0.09}_{-0.08}.
\end{equation}
Furthermore, LEP reports a 2.3$\sigma$ local excess in the $e^+e^-\to Z(H\to b\bar b)$ channel, which is compatible with a scalar particle of approximately 95 GeV \cite{LEPWorkingGroupforHiggsbosonsearches:2003ing}. The corresponding signal strength is \begin{equation}
\mu_{b\bar b}^{exp}(95) = 0.117 \pm 0.057.
\end{equation}

From a theoretical perspective, various BSM frameworks have been proposed to account for the observed excesses near 95 GeV~\cite{Cao:2016uwt,Cao:2019ofo,Aguilar-Saavedra:2023tql,Ahriche:2023hho,Ahriche:2023wkj}. Previous studies~\cite{n1,n2} showed that the diphoton rate can be several times larger than the SM prediction for a scalar of the same mass in the Next-to-Minimal Supersymmetric Standard Model (NMSSM). Similarly, the Two-Higgs-Doublet Model extended with an additional real singlet (N2HDM) has been extensively explored as a possible explanation for the CMS and LEP excesses~\cite{12,13}. Other scenarios, such as the $\mu\nu$SSM with CP conservation or violation~\cite{113,114,115}, singlet-extended or radiative Higgs models~\cite{16}, and gauge-extended frameworks like the $U(1)_X$ model~\cite{u1x}, have also been investigated as viable possibilities.

Despite the success of the SM in describing particle interactions, it still leaves several fundamental questions unanswered, such as the nature of dark matter, the origin of neutrino masses, and the hierarchy problem. These open issues have long motivated the exploration of BSM physics. Supersymmetry (SUSY), once regarded as one of the most elegant and theoretically appealing extensions, offers a natural framework to address these problems through its fermion-boson symmetry. However, the persistent null results in supersymmetric particle searches over the past few decades have substantially reduced its experimental viability. The reported 750 GeV excess in 2015 initially generated considerable excitement as a potential signal of NP, but the 2016 data did not confirm this excess~\cite{ATLAS:2016gzy,CMS:2016kgr}, further decreasing the overall interest in supersymmetric frameworks. Meanwhile, attention shifted toward dark matter, another compelling aspect of NP. Although significant progress has been made in dark matter searches, stringent direct-detection limits have excluded most of the viable parameter space \cite{Essig:2012yx,Clark:2020mna}, leaving only narrow regions open for exploration. Consequently, increasing interest has turned to non-supersymmetric extensions, which can offer simpler and more economical explanations. Specifically, the observation of a scalar lighter than 125 GeV would not necessarily point to a supersymmetric origin but could instead arise naturally in certain non-supersymmetric gauge extensions.

In the $U(1)_X$VLFM, the right-handed and left-handed fields of vector-like fermions have the same transformation properties under the gauge group of the SM, allowing them to acquire gauge-invariant mass terms \cite{Aguilar-Saavedra:2013qpa}. Consequently, they can evade the constraints from the Higgs production cross section and direct detection experiments at the LHC. The mixing between vector-like fermions and SM fermions induces corrections to the couplings of fermions with the $W$, $Z$, and Higgs bosons \cite{Cao:2022mif}. As a result, the Glashow-Iliopoulos-Maiani (GIM) mechanism is naturally violated, leading to flavor-changing neutral currents (FCNCs) at tree level. Moreover, vector-like fermions can introduce new CP-violating sources, contributing to the electric dipole moments (EDMs) of leptons, quarks, and neutrons \cite{Cao:2023smj}. Finally, the introduced vector-like quarks can help alleviate the gauge hierarchy problem in the SM. The $U(1)_X$VLFM has rich phenomenological research value.

The remainder of this paper is organized as follows. In Sec.II A, we introduce the main components of the $U(1)_X$VLFM, while Sec.II B presents the relevant formulas and mass matrices required for the one-loop corrections to the CP-even Higgs mass-squared matrix. In Sec.III, we derive the decay widths and corresponding signal strengths of the SM-like Higgs boson with a mass around 125 GeV, including the channels $h_{2} \to\gamma\gamma$, $h_{2} \to VV^*~(V=Z,W)$, and $h_{2} \to f\bar f~(f=b,\tau)$. Sec.IV is devoted to the analysis of the lighter Higgs-like scalar around 95 GeV, where we present its dominant decay modes $h_{1} \to\gamma\gamma$ and $h_{1} \to b\bar b$ together with the corresponding signal strengths. In Sec.V, we perform the numerical analysis based on the relevant parameter space, and Sec.VI summarizes our main results and discussions. For completeness, some lengthy analytical expressions are collected in Appendix \ref{A1}.
\section{The $U(1)_X$VLFM and the Higgs sector}

\subsection{The relevant content of $U(1)_X$VLFM}
The gauge group of the $U(1)_X$VLFM is $SU(3)_C \otimes SU(2)_L \otimes U(1)_Y \otimes U(1)_X$.
To construct the $U(1)_X$VLFM, the field content of the SM is extended by three generations of right-handed neutrino fields $\nu_R$, two singlet Higgs fields $\phi$ and $S$, and one generation of vector-like quark, lepton, and neutrino fields.
The singlet Higgs fields $\phi$ and $S$ are responsible for the spontaneous breaking of the additional $U(1)_X$ gauge symmetry and for generating the masses of the vector-like fermions.
The inclusion of right-handed neutrinos allows the light neutrino masses to be generated at the tree level via the seesaw mechanism.
After spontaneous symmetry breaking, the neutral CP-even components of $H,~\phi$, and $S$ mix with each other, forming a $3\times3$ mass-squared matrix in the scalar sector.
The loop corrections from the vector-like fermions can modify the scalar mass spectrum, allowing the model to accommodate both the SM-like Higgs boson around 125 GeV and a lighter scalar state near 95 GeV.
\begin{table}
\caption{Properties of new particles introduced in the model}
\begin{tabular}{|c|c|c|c|c|}
\hline
Field & $SU(3)_C$ & $SU(2)_L$ & $U(1)_Y$ & $U(1)_X$ \\
\hline
$\phi$ & 1 & 1 & 0 & $Q_a+Q_b$ \\
\hline
$S$ & 1 & 1 & 0 & $Q_a$ \\
\hline
$\nu_R$ & 1 & 1 & 0 & 0  \\
\hline
$d_{XL}$ & 3 & 1 & -1/3 & $Q_a$  \\
\hline
$u_{XL}$ & 3 & 1 & 2/3 & -$Q_a$  \\
\hline
$d_{XR}$ & $\bar{3}$ & 1 & 1/3 & $Q_b$ \\
\hline
$u_{XR}$ & $\bar{3}$ & 1 & -2/3 & -$Q_b$\\
\hline
$e_{XL}$ & 1 & 1 & -1 & $Q_a$ \\
\hline
$\nu_{XL}$ & 1 & 1 & 0 & -$Q_a$ \\
\hline
$e_{XR}$ & 1 & 1 & 1 & $Q_b$\\
\hline
$\nu_{XR}$ & 1 & 1 & 0 & -$Q_b$ \\
\hline
\end{tabular}
\label{fields}
\end{table}

The scalar sector of the $U(1)_X$VLFM consists of one $SU(2)_L$ Higgs doublet $H$ and two singlet Higgs fields $\phi$ and $S$. Their explicit forms are shown in the follow,
\begin{eqnarray}
&&H=\left(\begin{array}{c}H^0\\H^-\end{array}\right),
~~~~~~
H^0={1\over\sqrt{2}}\Big(v+\phi_H+i\sigma_H\Big),
\nonumber\\
&&\phi={1\over\sqrt{2}}\Big(v_P+\phi_P+i\sigma_P\Big),~~~~~~
S={1\over\sqrt{2}}\Big(v_S+\phi_S+i\sigma_S\Big).
\end{eqnarray}
where $v$, $v_P$ and $v_S$ denote the vacuum expectation values (VEVs) of the Higgs superfields $H$, $\phi$ and $S$, respectively.

The two scalars play distinct roles in fermion mass generation: the VEV $v_P$ of $\phi$ mainly contributes to the masses of vector-like fermions, while $v_S$ of $S$ controls the mixing between third-generation SM fermions and vector-like fermions. In the neutrino sector, both $v_P$ and $v_S$ contribute to neutrino masses via a seesaw mechanism. Therefore, two $U(1)_X$ Higgs scalars are necessary to obtain a realistic fermion mass spectrum and mixing structure. Introducing two singlet scalars also expands the parameter space, providing sufficient flexibility to simultaneously satisfy the mass conditions for the 95 GeV and 125 GeV scalar states and achieve a good fit to experimental data.

Although not required for anomaly cancellation, right-handed neutrinos are included to generate neutrino masses. Their contribution to the 95 GeV scalar signal is minor, entering only via loop corrections to the scalar effective potential. Hence, their primary role remains to provide a viable mechanism for light neutrino masses.

The scalar potential and Yukawa interactions relevant to the Higgs and vector-like fermion sectors can be expressed as
\begin{eqnarray}
&&\mathcal{L}=-\mu^2_H H^\dagger H-\mu^2_P |\phi|^2 -\mu^2_X |S|^2+\lambda_H(H^\dagger H)^2+\lambda_P |\phi|^4 +\lambda_X |S|^4\nonumber\\&&
~~+\lambda_{HP}(H^\dagger H)|\phi|^2+\lambda_{HX}(H^\dagger H)|S|^2+\lambda_{PX}|S|^2|\phi|^2\nonumber\\&&
~~-S d^*_{XL,k}Y^*_{XD,jk}d_{R,j}-S u^*_{R,j}Y_{XU,jk}u_{XL,k}-S e^*_{XL,k}Y^*_{XE,jk}e_{R,j}\nonumber\\&&
~~-S \nu^*_{R,j}Y_{XN,jk}\nu_{XL,k}-h.c.\nonumber\\&&
~~-\phi d^*_{XL,k}Y^*_{PD,jk}d_{XR,j}-\phi u^*_{XR,j}Y_{PU,jk}u_{XL,k}-\phi e^*_{XL,k}Y^*_{PE,jk}e_{XR,j}\nonumber\\&&
~~-\phi \nu^*_{XR,j}Y_{PN,jk}\nu_{XL,k}-h.c.\nonumber\\&&
~~-Y^*_{u,jk}\bar q_{L,k}H u_{R,j}+Y^*_{d,jk}\bar q_{L,k}\tilde{H} d_{R,j}+Y^*_{e,jk}\bar l_{k}\tilde{H} e_{R,j}+h.c.
\end{eqnarray}

The field content beyond SM and charge assignments of the $U(1)_X$VLFM are presented in Table~\ref{fields}, where $Y^Y$ and $Y^X$ denote the $U(1)_Y$ and $U(1)_X$ gauge charges, respectively. According to the textbook~\cite{Peskin}, the SM is free of gauge anomalies. In the extended $U(1)_X$VLFM, one generation of vector-like fermions is introduced precisely to preserve this anomaly-free structure.
The corresponding anomaly cancellation conditions can be verified as follows:

1. The pure gauge anomalies involving three $SU(3)_C$ or three $SU(2)_L$ gauge bosons vanish exactly as in the SM.

2. Mixed anomalies that contain one $SU(3)_C$ or one $SU(2)_L$ gauge boson are proportional to the traces $Tr[t^a]=0$ or $Tr[\tau^a]=0$, and hence they also vanish.

3. The mixed anomalies with one $U(1)_Y$ or $U(1)_X$ gauge boson and two $SU(3)_C$ bosons are proportional to the group-theoretical factors
$Tr[t^at^bY^Y]=\frac{1}{2}\delta^{ab}\sum_q Y^Y_q$ or $Tr[t^at^bY^X]=\frac{1}{2}\delta^{ab}\sum_q Y^X_q$.

4. Similarly, the mixed anomalies involving one $U(1)_Y$ or $U(1)_X$ gauge boson and two $SU(2)_L$ bosons are proportional to
$Tr[\tau^a\tau^bY^Y]=\frac{1}{2}\delta^{ab}\sum_{L} Y^Y_{L}$ or $Tr[\tau^a\tau^bY^X]=\frac{1}{2}\delta^{ab}\sum_{L} Y^X_{L}$.

5. The pure Abelian anomalies containing three $U(1)$ gauge bosons can be categorized into four types:
\begin{eqnarray}
&&Tr[Y^YY^YY^Y]=\sum_n(Y^Y_n)^3,~~~~~~~~~~~Tr[Y^XY^XY^X]=\sum_n(Y^X_n)^3,\nonumber\\&&
Tr[Y^XY^YY^Y]=\sum_nY^X_n(Y^Y_n)^2,~~~~~~Tr[Y^YY^XY^X]=\sum_nY^Y_n(Y^X_n)^2.
\end{eqnarray}

6. The mixed gravitational anomalies with one $U(1)$ gauge boson are proportional to
$Tr[Y^Y]=\sum_n Y_n^Y$ or $Tr[Y^X]=\sum_n Y_n^X$.

The anomaly structures that do not involve the $U(1)_X$ gauge group are identical to those in the SM and remain automatically consistent. For the anomaly terms associated with the additional $U(1)_X$ gauge symmetry, the inclusion of one generation of vector-like fermions guarantees their complete cancellation. Therefore, all gauge and mixed gravitational anomalies are successfully canceled in the $U(1)_X$VLFM, confirming the theoretical consistency of the model.

The coexistence of two Abelian gauge groups $U(1)_Y$ and $U(1)_X$ in the $U(1)_X$VLFM introduces a new feature absent in the SM with a single $U(1)_Y$: the gauge kinetic mixing. Even if this mixing is set to zero at $M_{GUT}$, it can be radiatively generated through RGEs.

The covariant derivative of this model takes the general form
\begin{eqnarray}
&&D_\mu=\partial_\mu-i\left(\begin{array}{cc}Y^Y,&Y^X\end{array}\right)
\left(\begin{array}{cc}g_{Y},&g{'}_{{YX}}\\g{'}_{{XY}},&g{'}_{{X}}\end{array}\right)
\left(\begin{array}{c}A_{\mu}^{\prime Y} \\ A_{\mu}^{\prime X}\end{array}\right)\;,
\label{gauge1}
\end{eqnarray}
we redefine the gauge fields as
\begin{eqnarray}
&&
B_\mu = g_Y A_\mu^{'Y} + g'_{YX} A_\mu^{'X}, \quad
B'_\mu = g'_{XY} A_\mu^{'Y} + g'_X A_\mu^{'X}.
\end{eqnarray}

The gauge fields $B_\mu$ and $B_\mu'$ have canonical kinetic terms
\begin{eqnarray}
&&
-\frac{1}{4} B_{\mu\nu} B^{\mu\nu} - \frac{1}{4} B'_{\mu\nu} B^{\prime\mu\nu}.\nonumber\\
&&B_{\mu \nu} = \partial_\mu B_\nu - \partial_\nu B_\mu, \quad B'_{\mu \nu} = \partial'_\mu B_\nu - \partial'_\nu B_\mu.
\end{eqnarray}

Substituting these redefinitions into the kinetic terms and expanding, one obtains the standard form
\begin{eqnarray}
&&
\mathcal{L}_{\text{kin}} = -\frac{1}{4} F_{\mu\nu} F^{\mu\nu}
- \frac{1}{4} F'_{\mu\nu} F'^{\mu\nu}
- \frac{\epsilon}{4} F_{\mu\nu} F'^{\mu\nu}\nonumber\\
&&F_{\mu \nu} = \partial_\mu A_\nu^{'Y} - \partial_\nu A_\mu^{'Y}, \quad F'_{\mu \nu} = \partial_\mu A_\nu^{'X} - \partial_\nu A_\mu^{'X}.
\end{eqnarray}
where the kinetic mixing parameter $\epsilon$ is given by
\begin{eqnarray}
&&
\epsilon =
\frac{2\bigl(g'_X g'_{YX} + g'_{XY} g_Y\bigr)}
{\sqrt{\bigl(g_X'^2 + g_{XY}'^2 \bigr)\bigl(g_Y^2 + g_{YX}'^2 \bigr)}}.
\end{eqnarray}

This expression provides the explicit relation between $g'_{YX}$, $g'_{XY}$ and the kinetic mixing parameter $\epsilon$.

$A_{\mu}^{\prime Y}$ and $A_{\mu}^{\prime X}$ denote the gauge fields associated with $U(1)_Y$ and $U(1)_X$, while $Y^Y$ and $Y^X$ are the corresponding charges. Since both Abelian symmetries remain unbroken prior to symmetry breaking, a proper basis rotation can be performed in the gauge field space. Introducing a transformation matrix $R$, the gauge coupling matrix can be brought into an upper triangular form:
\begin{eqnarray}
&&\left(\begin{array}{cc}g_{Y},&g{'}_{{YX}}\\g{'}_{{XY}},&g{'}_{{X}}\end{array}\right)
R^T=\left(\begin{array}{cc}g_{1},&g_{{YX}}\\0,&g_{{X}}\end{array}\right)\;.
\label{gauge2}
\end{eqnarray}

Consequently, the $U(1)$ gauge fields are redefined as
\begin{eqnarray}
&&R\left(\begin{array}{c}A_{\mu}^{\prime Y} \\ A_{\mu}^{\prime X}\end{array}\right)
=\left(\begin{array}{c}A_{\mu}^{Y} \\ A_{\mu}^{X}\end{array}\right)\;.
\label{gauge3}
\end{eqnarray}

At the tree level, the neutral gauge bosons $A^Y_\mu$, $V^3_\mu$, and $A^X_\mu$ mix with each other, leading to a non-diagonal mass matrix in the basis $(A^Y_\mu, V^3_\mu, A^X_\mu)$, which can be written as
\begin{eqnarray}
&&\left(\begin{array}{*{20}{c}}
\frac{1}{4}g_{1}^2 v^2 &~~ -\frac{1}{4}g_{1}g_{2} v^2 & ~~\frac{1}{4}g_{1}g_{{YX}} v^2 \\
-\frac{1}{4}g_{1}g_{2} v^2 &~~ \frac{1}{4}g_{2}^2 v^2 & ~~-\frac{1}{4}g_{2}g_{{YX}} v^2\\
\frac{1}{4}g_{1}g_{{YX}} v^2 &~~ -\frac{1}{4}g_{2}g_{{YX}} v^2 &~~\frac{1}{4}g_{{YX}}^2 v^2+\frac{1}{4}g_{{X}}^2\xi^2
\end{array}\right)\label{gauge matrix}
\end{eqnarray}
with $\xi^2=4(Q_a + Q_b)^2 v^2_P+ 4Q^2_a v^2_S$. The diagonalization of the above matrix can be achieved through a rotation involving the weak mixing angle $\theta_W$ and the additional angle $\theta_W'$,
\begin{eqnarray}
&&\left(\begin{array}{*{20}{c}}
\gamma_\mu\\ [6pt]
Z_\mu\\ [6pt]
Z'_\mu
\end{array}\right)=
\left(\begin{array}{*{20}{c}}
\cos\theta_{W} & \sin\theta_{W} & 0 \\ [6pt]
-\sin\theta_{W}\cos\theta_{W}' & \cos\theta_{W}\cos\theta_{W}' & \sin\theta_{W}'\\ [6pt]
\sin\theta_{W}\sin\theta_{W}' & -\cos\theta_{W}'\sin\theta_{W}' & \cos\theta_{W}'
\end{array}\right)
\left(\begin{array}{*{20}{c}}
A^Y_\mu\\ [6pt]
V^3_\mu\\ [6pt]
A^{X}_\mu
\end{array}\right).
\end{eqnarray}

The explicit expression of $\sin^2\theta_W'$ is given by
\begin{eqnarray}
\sin^2\theta_{W}'=\frac{1}{2}-\frac{(g_{{YX}}^2-g_{1}^2-g_{2}^2)v^2+
g_{X}^2\xi^2}{2\sqrt{(g_{{YX}}^2+g_{1}^2+g_{2}^2)^2v^4+2g_{X}^2(g_{{YX}}^2-g_{1}^2-g_{2}^2)v^2\xi^2+g_{X}^4\xi^4}}.
\end{eqnarray}
where the order of magnitude of $\sin\theta_W'$ is about $\mathcal{O}(10^{-3})$~\cite{Barate:1999qx,Abreu:2000ap,King:2005jy}.

The new mixing angle $\theta_{W}^\prime$ appears in the couplings involving $Z$ and $Z^{\prime}$. The exact eigenvalues of Eq. (\ref{gauge matrix}) are calculated
\begin{eqnarray}
&&\qquad\;\quad\;m_\gamma^2=0,\nonumber\\
&&\qquad\;\quad\;m_{Z,{Z^{'}}}^2=\frac{1}{8}\Big((g_{1}^2+g_2^2+g_{YX}^2)v^2+g_{X}^2 \xi^2 \nonumber\\
&&\qquad\;\qquad\;\qquad\;\mp\sqrt{[(g_{1}^2+g_2^2+g_{YX}^2)v^2+g_{X}^2 \xi^2]^2-4(g^2_1+g^2_2)g^2_X v^2 \xi^2}\Big).
\label{mz}
\end{eqnarray}

In our model, due to the mixing between the $Z$ boson and the TeV-scale $Z'$, the $Z$ boson mass does receive corrections. Expanding the exact expression (Eq. \ref{mz}) in the limit $v^2 \ll \xi^2$, we obtain:
\begin{eqnarray}
&&m^2_Z\approx \frac{(g_1^2 + g_2^2)v^2}{4}
- \frac{(g_1^2 + g_2^2) g_{YX}^2}{4 g_X^2} \cdot \frac{v^4}{\xi^2},
\end{eqnarray}
where $\xi^2 = 4(Q_a+Q_b)^2 v_P^2 + 4Q_a^2 v_S^2$, and we choose $Q_a=1,~Q_b=1$.

The leading term reproduces the SM result, and a further estimate gives
\begin{eqnarray}
&&
\frac{\delta m^2_Z}{m^2_Z} \sim \frac{g_{YX}^2}{g_X^2} \cdot \frac{v^2}{\xi^2} =\frac{g_{YX}^2}{g_X^2} \cdot \frac{v^2}{16 v_P^2 + 4 v_S^2} \sim\mathcal{O}(10^{-5}).
\end{eqnarray}

For the $W$ boson mass, since the new scalar fields are singlets under $SU(2)_L$ and no charged scalar states (such as $H^{\pm}$) are present, there is no mixing involving the $W$ boson. Moreover, electroweak symmetry breaking is still dominated by the SM Higgs doublet, so scalar mixing does not introduce additional corrections to the $W$ mass. Its tree-level mass expression remains
\begin{eqnarray}
&&
m_W^2 = \frac{g_2^2 v^2}{4},
\end{eqnarray}
and is not affected by the above mixing effects in this model.

Based on the above analysis, within the parameter regions considered in Figs.~\ref{tu4} and~\ref{tu5}, the corrections to $m_Z$ and $m_W$ are negligible, and thus they are not included in the fit.

\subsection{Higgs mass correction}
The one-loop effective potential can be written in the following form
\begin{eqnarray}
&&V_{Total}=V_{Tree}+\Delta V.
\end{eqnarray}

Here, $V_{Tree}$ denotes the tree-level potential, while $\Delta V$ corresponds to the one-loop correction.

The simplified Higgs potential at tree level is given below
\begin{eqnarray}
&&V_{Tree}=\mu^2_H H^\dagger H+\mu^2_P |\phi|^2 +\mu^2_X |S|^2-\lambda_H(H^\dagger H)^2-\lambda_P |\phi|^4 -\lambda_X |S|^4\nonumber\\&&
~~~~~~~-\lambda_{HP}(H^\dagger H)|\phi|^2-\lambda_{HX}(H^\dagger H)|S|^2-\lambda_{PX}|S|^2|\phi|^2.
\end{eqnarray}

We also derive the corresponding tadpole equations at tree level
\begin{eqnarray}
&&2\lambda_H v^2-2\mu^2_H +\lambda_{HP} v^2_P+\lambda_{HX} v^2_S=0, \\
&&2\lambda_X v^2_S-2\mu^2_X +\lambda_{HX} v^2+\lambda_{PX} v^2_P=0,\\
&&2\lambda_P v^2_P-2\mu^2_P +\lambda_{HP} v^2+\lambda_{PX} v^2_S=0.
\end{eqnarray}

In addition, the tree-level mass-squared matrix for the CP-even Higgs $({\phi}_{H}, {\phi}_{S}, {\phi}_{P})$ is presented
\begin{eqnarray}
M^2_{h,tree} = \left(
\begin{array}{ccc}
m_{{\phi}_{H}{\phi}_{H}} &-\lambda_{HX}v v_S &-\lambda_{HP}v v_P \\
-\lambda_{HX}v v_S &m_{{\phi}_{S}{\phi}_{S}} &-\lambda_{PX}v_P v_S \\
-\lambda_{HP}v v_P &-\lambda_{PX}v_P v_S &m_{{\phi}_{P}{\phi}_{P}} \end{array}
\right),
\end{eqnarray}
with the diagonal entries defined as
\begin{eqnarray}
&&m_{\phi_{H}\phi_{H}}=\frac{1}{2}\Big(-6\lambda_H v^2-\lambda_{HP}v^2_P-\lambda_{HX}v^2_S\Big)+\mu^2_H,
\\&&m_{\phi_{S}\phi_{S}} = \frac{1}{2}\Big(-6\lambda_X v^2_S-\lambda_{HX}v^2-\lambda_{PX}v^2_P\Big)+\mu^2_X,
\\ &&m_{\phi_{P}\phi_{P}} = \frac{1}{2}\Big(-6\lambda_P v_P^2-\lambda_{HP}v^2-\lambda_{PX}v^2_S\Big)+\mu^2_P.
\end{eqnarray}

Within dimensional regularization and the $\overline{\text{MS}}$ renormalization scheme in the Landau gauge, the effective Higgs potential receives one-loop corrections, with $\Delta V$ given in the Coleman-Weinberg form \cite{ZSM1,ZSM2,ZSM3}
\begin{eqnarray}
&&\Delta V=\sum_i\frac{n_i}{64\pi^2}m_i^4(\phi_H,\phi_S,\phi_P)\Big(
\log\frac{m_i^2(\phi_H,\phi_S,\phi_P)}{Q^2}-\frac{3}{2}\Big).
\end{eqnarray}

The renormalization scale $Q$ is chosen to be of the order of TeV. The degrees of freedom $n_i$ for each mass eigenstate are assigned as follows: -12 for quarks, -4 for charged leptons, -2 for neutrinos, 6 for the $W$ boson, 3 for the $Z$ boson, and similarly for the $Z^\prime$ boson. The specific potential for the one-loop correction reads
\begin{eqnarray}
&&\Delta V = \sum_{f=t,b,\tau} V_f + \sum_{F=t',b',\tau'} V_F +V_{\nu'}+ V_W + V_Z + V_{Z'}.
\end{eqnarray}

Here, $V_f$ ($f=t,b,\tau$) and $V_F$ ($F=t',b',\tau'$) denote the one-loop effective potential contributions from the SM fermions
and the vector-like fermions, respectively. $V_{\nu'}$ represents the contributions from the heavy neutrino sector. $V_W$, $V_Z$, and $V_{Z'}$ correspond to the contributions from the $W$ boson, $Z$ boson, and the $Z^\prime$ boson.

In the $U(1)_X$VLFM, the fermion mass matrices arise from the Yukawa interactions and are defined in the gauge eigenstate basis. For instance, the down-type quarks are written in the basis $(d_L, d_{XL})$ and $(d^*_R, d^*_{XR})$, leading to the mass matrix
\begin{equation}
m_d = \left(
\begin{array}{cc}
\frac{1}{\sqrt{2}}v Y^T_d &0\\
\frac{1}{\sqrt{2}}v_S Y^T_{XD}  &\frac{1}{\sqrt{2}}v_P Y^T_{PD}\end{array}
\right).
\label{d}
\end{equation}

Similarly, the up-type quarks and charged leptons are expressed in the corresponding bases $(u_L, u_{XL}), (u^*_R, u^*_{XR})$ and $(e_L, e_{XL}), (e^*_R, e^*_{XR})$, with mass matrices
\begin{eqnarray}
&&\hspace{1cm}
m_u = \left(
\begin{array}{cc}
\frac{1}{\sqrt{2}}v Y^T_u &0\\
\frac{1}{\sqrt{2}}v_S Y^T_{XU}  &\frac{1}{\sqrt{2}}v_P Y^T_{PU}\end{array}\right),~~
m_e = \left(
\begin{array}{cc}
\frac{1}{\sqrt{2}}v Y^T_e &0\\
\frac{1}{\sqrt{2}}v_S Y^T_{XE}  &\frac{1}{\sqrt{2}}v_P Y^T_{PE}\end{array}
\right).
\label{ue}
\end{eqnarray}

These Dirac-type mass matrices are diagonalized by bi-unitary transformations,
\begin{equation}
U_L^{f,*}\, m_f \, U_R^{f,\dagger} = m_f^{\text{diag}}, \qquad f = d,u,e,
\end{equation}
where $U_L^f$ and $U_R^f$ are unitary matrices acting on the left- and right-handed fermion fields, respectively.

The neutrino sector, in the basis $(\nu_L, \nu^*_R, \nu_{XL}, \nu^*_{XR})$, has the mass matrix
\begin{equation}
m_{\nu} = \left(
\begin{array}{cccc}
0 &\frac{1}{\sqrt 2}v Y^T_v & 0 &0  \\
\frac{1}{\sqrt 2}v Y_v &0& \frac{1}{\sqrt 2}v_S Y_{XN} &0 \\
0 &\frac{1}{\sqrt 2}v_S Y^T_{XN}& 0 &\frac{1}{\sqrt 2}v_P Y^T_{PN}\\
0 &0& \frac{1}{\sqrt 2}v_P Y_{PN} &0\end{array}
\right).
 \end{equation}

Being Majorana-type, the neutrino mass matrix is diagonalized by a single unitary matrix $U^V$ as
\begin{equation}
U^{V,*} m_\nu U^{V,\dagger} = m_\nu^{\text{diag}}.
\end{equation}

The one-loop minimization conditions of the total effective potential $V_{Total}$ in the $U(1)_X$VLFM are given by
\begin{eqnarray}
&&\left \langle \frac{\partial V_{Total}}{\partial \phi_H} \right \rangle=
\left \langle \frac{\partial V_{Total}}{\partial \phi_S} \right \rangle=
\left \langle \frac{\partial V_{Total}}{\partial \phi_P} \right \rangle=0,
\end{eqnarray}

Since the analytic forms of these conditions are very tedious, we solve them numerically. The explicit expressions for the down-type quark sector are provided in Appendix \ref{A1} for reference.

The one-loop contributions to the effective potential $V_{Total}$ induce radiative corrections to the mass-squared matrix of the CP-even Higgs sector
\begin{eqnarray}
M^2_{Total,h}=M^2_{Tree,h}+\Delta M^2_h,
\end{eqnarray}

The elements of the corrected mass-squared matrix $M^2_{Total,h,ij}$ are derived from the second derivatives of the effective potential $V_{{Total}}$
\begin{eqnarray}
 M^2_{Total,h,ij}=\Big\langle\frac{\partial^2V_{Total}}{\partial \phi_i \partial \phi_j}\Big|_{\phi_i,\phi_j=\phi_H,\phi_S,\phi_P}\Big\rangle.
\end{eqnarray}

Diagonalization of $M^2_{Total,h}$ yields the square of the mass eigenvalues $m_{{h_n}}$ ($n=1,2,3$), which we order as $m_{{h_1}} < m_{{h_2}} < m_{{h_3}}$. In this work, we identify the lightest eigenstate $h_1$ with the Higgs-like excess observed near 95 GeV, and the next-to-lightest eigenstate $h_2$ with the SM-like Higgs boson at 125 GeV.
\section{The 125 GeV Higgs decays}
At the LHC, the dominant production channel of the Higgs boson is gluon fusion. In the SM, this process is induced at leading order (LO) by a one-loop diagram involving virtual top quarks, while subleading contributions from bottom quarks are strongly suppressed. The inclusive cross section has been calculated up to next-to-next-to-leading order (NNLO) in QCD~\cite{NNLO}, which enhances the LO prediction by about 80-100\%. Since the gluon fusion rate is loop-induced, it is highly sensitive to NP: any heavy particle that couples significantly to the Higgs can modify the amplitude and hence the production rate. In the $U(1)_X$VLFM, the LO decay width for the process $h_2\rightarrow gg$ is given by \cite{Gamma1,Gamma2,Gamma3,Gamma4,Gamma5,Gamma6}
\begin{eqnarray}
&&\Gamma_{_{NP}}(h_2\rightarrow gg)={G_{_F}\alpha_s^2m_{_{h_2}}^3\over64\sqrt{2}\pi^3}
\Big|\sum\limits_{q=t,b,t',b'}g_{_{h_2qq}}A_{1/2}(x_q)\Big|^2,
\label{hgg}
\end{eqnarray}
where $x_a = m^2_{h_{n}} / (4m_a^2)$. The concrete expressions for the Yukawa couplings $g_{h u u}$ and $g_{h d d}$ are formulated as
\begin{eqnarray}
&&g_{h_{n} u_j u_j}=- \frac{v}{m_{u_j}}
\Big[-\frac{1}{\sqrt 2} \Big(\sum_{a,b=1}^3Y^*_{u, a b}U^u_{R,ja}U^u_{L,jb}Z^H_{n1}\nonumber\\
&&~~~~~~~~~+\sum_{a}^3Y^*_{XU,a 1}U^u_{R,ja}U^u_{L,j4}Z^H_{n2}+Y^*_{PU}U^u_{R,j4}
U^u_{L,j4}Z^H_{n3}\Big)\Big],\quad (j=1,\dots,4),   \nonumber\\
&&g_{h_{n} d_j d_j}=-\frac{v}{m_{d_j}}\Big[-\frac{1}{\sqrt 2}\Big(\sum_{a,b=1}^3Y^*_{d,ab}U^d_{R,ja}U^d_{L,jb}Z^H_{n1}\nonumber\\
&&~~~~~~~~~+\sum_{a=1}^3Y^*_{XD,a1}U^d_{R,ja}U^d_{L,j4}Z^H_{n2}+Y^*_{PD}U^d_{R,j4}U^d_{L,j4}Z^H_{n3}\Big)\Big],\quad (j=1,\dots,4),
\end{eqnarray}
with
\begin{eqnarray}
&&Y^*_{XU,a 1} = \left(
\begin{array}{c}
0\\
0\\
Y_{XU}\end{array}
\right)~~~~\text{and}~~~~~
Y^*_{XD,a 1} = \left(
\begin{array}{c}
0\\
0\\
Y_{XD}\end{array}
\right).
\end{eqnarray}

Here $j=1,2,3$ correspond to the SM quarks $(u,c,t)$ and $(d,s,b)$, while $j=4$ denotes the vector-like quarks $(t',b')$ introduced in the $U(1)_X$VLFM extension. $Z^H$ is the Higgs mixing matrix. The Yukawa couplings $Y_{XU,XD}$ and $Y_{PU,PD}$ describe the interactions of the Higgs with the vector-like fermions, leading to mixing with the SM quarks.

The form factor $A_{1/2}$ in Eq. (\ref{hgg}) is defined as
\begin{eqnarray}
&&A_{1/2}(x)=2\Big[x+(x-1)g(x)\Big]/x^2,
\end{eqnarray}
with
\begin{eqnarray}
&&g(x)=\left\{\begin{array}{l}\arcsin^2\sqrt{x},\;x\le1\\
-{1\over4}\Big[\ln{1+\sqrt{1-1/x}\over1-\sqrt{1-1/x}}-i\pi\Big]^2,\;x>1\;.\end{array}\right.
\end{eqnarray}

The Higgs diphoton decay is also mediated by loop diagrams. In the SM, the LO contributions arise from virtual charged gauge bosons $W^\pm$ or virtual top quarks. Within the $U(1)_X$VLFM framework, additional effects are introduced by the vector-like
 fermions $t'$, $b'$ and $\tau'$ mixing respectively with the third-generation fermions. The decay width can be written as
\begin{eqnarray}
&&\Gamma_{_{NP}}(h_2\rightarrow\gamma\gamma)={G_{_F}\alpha^2m_{_{h_2}}^3\over128\sqrt{2}\pi^3}
\Big|\sum\limits_fN_cQ_{f}^2g_{_{h_2ff}}A_{1/2}(x_f)+g_{_{h_2WW}}A_1(x_{_{\rm W}})
\Big|^2\;,
\label{hrr}
\end{eqnarray}
where the sum runs over the charged fermions $f = t, b, \tau, t', b', \tau'$. Here $N_c$ denotes the color factor ($N_c=3$ for quarks, $N_c=1$ for leptons), and $Q_f$ is the corresponding electric charge.

The couplings of the Higgs to fermions $g_{h_{n} f f}$ are obtained from the diagonalization of the extended fermion mass matrices. For the leptonic sector, the coupling to mass eigenstate $l_j$ is
\begin{eqnarray}
&&g_{h_{n} l_j l_j}=-\frac{v}{m_{l_j}}\Big[-\frac{1}{\sqrt 2}\Big(\sum_{a,b=1}^3 Y^*_{e, a b}U^e_{R,ja}U^e_{L,jb}Z^H_{n1}\nonumber\\
&&~~~~~~~~+\sum_{a}^3 Y^*_{XE,a 1}U^e_{R,ja}U^e_{L,j4}Z^H_{n2}+Y^*_{PE}U^e_{R,j4}U^e_{L,j4}Z^H_{n3}\Big)\Big],\quad (j=1,\dots,4),
\end{eqnarray}
with
\begin{eqnarray}
&&Y^*_{XE,a 1} = \left(
\begin{array}{c}
0\\
0\\
Y_{XE}\end{array}
\right).
\end{eqnarray}

The Higgs coupling to $W$ bosons is determined by the Higgs mixing matrix
\begin{eqnarray}
&&g_{h_{n} W W}=Z^H_{n1}\;.
\end{eqnarray}

The loop function for spin-1 particles is given by
\begin{eqnarray}
&&A_1(x)=-\Big[2x^2+3x+3(2x-1)g(x)\Big]/x^2\;.
\end{eqnarray}

For the neutral CP-even Higgs boson with a mass of approximately $125~{\rm GeV}$, the decay channels $h_{2} \rightarrow ZZ^*$ and $h_{2} \rightarrow WW^*$ are allowed. Summing over all kinematically accessible final states of these off-shell bosons, the corresponding partial decay widths can be expressed as \cite{Keung1,Keung2,HtoVV-SUSY1,HtoVV-SUSY2}
\begin{eqnarray}
&&\Gamma(h_{2} \rightarrow WW^*)={3e^4m_{_{h_{2} }}\over512\pi^3s_{_{\rm W}}^4}|g_{_{h_2WW}}|^2
F({m_{_{\rm W}}\over m_{_{h_{2} }}}),\;\nonumber\\
&&\Gamma(h_{2} \rightarrow ZZ^*)={e^4m_{_{h_{2}}}\over2048\pi^3s_{_{\rm W}}^4c_{_{\rm W}}^4}|g_{_{h_{2} ZZ}}|^2
\Big(7-{40\over3}s_{_{\rm W}}^2+{160\over9}s_{_{\rm W}}^4\Big)F({m_{_{\rm Z}}\over m_{_{h_{2} }}}).\;
\end{eqnarray}
Here, we adopt the abbreviations $c_{_{\rm W}}=\cos\theta_{_{\rm W}}$ and $s_{_{\rm W}}=\sin\theta_{_{\rm W}}$ with $\theta_{_{\rm W}}$ denoting the Weinberg angle. Furthermore, $e$ is the electromagnetic coupling constant.

The $\rm Higgs-Z-Z$ coupling is
\begin{eqnarray}
&&g_{h_{n}  Z Z}=\frac{v}{2m^2_{Z}}\Big[\frac{1}{2}\Big(v\Big(g_1\cos\theta'_W\sin\theta_W+g_2\cos\theta_W\cos\theta'_W-g_{YX}\sin\theta'_W\Big)^2Z^H_{n1}\nonumber\\
&&~~~~~~~~+\Big(-2g_X\sin\theta'_W\Big)^2 \Big(Q_a^2 v_S Z^H_{n2}+\Big(Q_a+Q_b\Big)^2v_P Z^H_{n3}\Big)\Big)\Big],
\end{eqnarray}
and the form factor $F(x)$ is given as
\begin{eqnarray}
&&F(x)=-(1-x^2)\Big({47\over2}x^2-{13\over2}+{1\over x^2}\Big)-3(1-6x^2+4x^4)\ln x
\nonumber\\
&&\hspace{1.5cm}
+{3(1-8x^2+20x^4)\over\sqrt{4x^2-1}}\cos^{-1}\Big({3x^2-1\over2x^3}\Big)\;.\nonumber\\
\label{form-factor1}
\end{eqnarray}

The partial decay width of the neutral CP-even Higgs boson with a mass of $125~{\rm GeV}$ into a pair of fermions can be expressed at LO as~\cite{hff1,hff2}
\begin{eqnarray}
&&\Gamma_{{\rm{NP}}}(h_2 \rightarrow f\bar{f})=N_c {G_{F} m_{f}^2 m_{h_{2} } \over 4\sqrt{2}\pi}|g_{{h_{2} ff}}|^2 (1-\frac{4m_f^2}{m^2_{h_{2} }})^{3/2} \quad (f=b,\tau).
\label{hff}
\end{eqnarray}

Both ATLAS and CMS have reported a mild excess in Higgs production and decay into the diphoton channel relative to the SM expectation. The signal strength for a given production and decay channel, normalized to the corresponding SM prediction, is defined as~\cite{ratios}
\begin{eqnarray}
&&\mu_{\gamma\gamma,VV^*}^{{\rm{ggF}}}(125)={ \sigma_{{\rm{NP}}}({\rm{ggF}})\over
\sigma_{{\rm{SM}}}({\rm{ggF}})}\:{{\rm{BR}}_{{\rm{NP}}}(h_{2} \rightarrow\gamma\gamma,VV^*)\over
{\rm{BR}}_{{\rm{SM}}}(h_{2} \rightarrow\gamma\gamma,VV^*)}   \qquad (V=Z,W), \nonumber\\
&&\mu_{f\bar{f}}^{{\rm{VBF}}}(125)={ \sigma_{{\rm{NP}}}({\rm{VBF}})\over
\sigma_{{\rm{SM}}}({\rm{VBF}})}\:{{\rm{BR}}_{{\rm{NP}}}(h_{2} \rightarrow{f\bar{f}})\over
{\rm{BR}}_{{\rm{SM}}}(h_{2} \rightarrow{f\bar{f}})}  \qquad (f=b,\tau),
\label{eq-ratios}
\end{eqnarray}
where ggF and VBF stand for gluon-gluon fusion and vector boson fusion, respectively. Normalized to the SM values, the NP-to-SM ratios of Higgs production cross sections are determined by the ratios of the relevant partial and total decay widths
\begin{eqnarray}
&&{\sigma_{{\rm{NP}}}({\rm{ggF}})\over
\sigma_{{\rm{SM}}}({\rm{ggF}})} \approx {\Gamma_{{\rm{NP}}}(h_{2} \rightarrow gg) \over
\Gamma_{{\rm{SM}}}(h_{2} \rightarrow gg)} = {\Gamma_{{\rm{NP}}}^{h_{2} }\over
\Gamma_{{\rm{SM}}}^{h_{2} }}\: {\Gamma_{{\rm{NP}}}(h_{2} \rightarrow gg)/\Gamma_{{\rm{NP}}}^{h_{2} } \over
\Gamma_{{\rm{SM}}}(h_{2} \rightarrow gg)/\Gamma_{{\rm{SM}}}^{h_{2} }}\nonumber\\
&&\qquad\qquad\;\;={\Gamma_{{\rm{NP}}}^{h_{2} }\over
\Gamma_{{\rm{SM}}}^{h_{2}}}\: {{\rm{BR}}_{{\rm{NP}}}(h_{2} \rightarrow gg)\over
{\rm{BR}}_{{\rm{SM}}}(h_{2} \rightarrow gg)},\nonumber\\
&&{\sigma_{{\rm{NP}}}({\rm{VBF}})\over
\sigma_{{\rm{SM}}}({\rm{VBF}})} \approx {\Gamma_{{\rm{NP}}}(h_{2} \rightarrow{VV^*}) \over
\Gamma_{{\rm{SM}}}(h_{2} \rightarrow{VV^*})} ={\Gamma_{{\rm{NP}}}^{h_{2} }\over
\Gamma_{{\rm{SM}}}^{h_{2}} }\: {\Gamma_{{\rm{NP}}}(h_{2} \rightarrow{VV^*})/\Gamma_{{\rm{NP}}}^{h_{2} } \over
\Gamma_{{\rm{SM}}}(h_{2} \rightarrow{VV^*})/\Gamma_{{\rm{SM}}}^{h_{2} }}\nonumber\\
&&\qquad\qquad\quad ={\Gamma_{{\rm{NP}}}^{h_{2} }\over
\Gamma_{{\rm{SM}}}^{h_{2}}}\: {{\rm{BR}}_{{\rm{NP}}}(h_{2} \rightarrow{VV^*})\over
{\rm{BR}}_{{\rm{SM}}}(h_{2} \rightarrow{VV^*})}.
\label{eq-cross}
\end{eqnarray}
Here the total decay width of the 125 GeV Higgs boson in the NP framework is given by
\begin{eqnarray}
&&\Gamma_{{\rm{NP}}}^{h_{2} }=\sum\limits_{f=b,\tau,c,s} \Gamma_{{\rm{NP}}}(h_{2} \rightarrow f\bar{f})+ \sum\limits_{V=Z,W} \Gamma_{{\rm{NP}}}(h_{2} \rightarrow VV^*) \nonumber\\
&&\qquad\quad +\: \Gamma_{{\rm{NP}}}(h_{2} \rightarrow gg) +\Gamma_{{\rm{NP}}}(h_{2} \rightarrow \gamma\gamma),
\end{eqnarray}
in which contributions from rare or invisible decay channels are neglected, and $\Gamma_{{\rm{SM}}}^{h_{2} }$ denotes the total decay width of the SM Higgs boson. Using Eqs.~(\ref{eq-ratios}) and (\ref{eq-cross}), the signal strengths for the Higgs decay channels in the $U(1)_X$VLFM can be expressed as
\begin{eqnarray}
&&\mu_{\gamma\gamma}^{{\rm{ggF}}}(125)\approx {\Gamma_{{\rm{NP}}}(h_{2} \rightarrow gg) \over
\Gamma_{{\rm{SM}}}(h_{2} \rightarrow gg)} \:{\Gamma_{{\rm{NP}}}(h_{2} \rightarrow\gamma\gamma)/\Gamma_{{\rm{NP}}}^{h_{2} }\over
\Gamma_{{\rm{SM}}}(h_{2} \rightarrow\gamma\gamma)/\Gamma_{{\rm{SM}}}^{h_{2}}} \nonumber\\
&&\qquad\;\, ~~~~~={\Gamma_{{\rm{SM}}}^{h_{2} }\over \Gamma_{{\rm{NP}}}^{h_{2} }}\:{\Gamma_{{\rm{NP}}}(h_{2} \rightarrow gg) \over
\Gamma_{{\rm{SM}}}(h_{2} \rightarrow gg)}\:{\Gamma_{{\rm{NP}}}(h_{2} \rightarrow\gamma\gamma)\over
\Gamma_{{\rm{SM}}}(h_{2} \rightarrow\gamma\gamma)}, \nonumber\\
&&\mu_{VV^*}^{{\rm{ggF}}}(125)\approx {\Gamma_{{\rm{NP}}}(h_{2} \rightarrow gg)\over
\Gamma_{{\rm{SM}}}(h_{2} \rightarrow gg)} \:{\Gamma_{{\rm{NP}}}(h_{2} \rightarrow VV^*) /\Gamma_{{\rm{NP}}}^{h_{2}} \over \Gamma_{{\rm{SM}}}(h_{2} \rightarrow VV^*) /\Gamma_{{\rm{SM}}}^{h_{2}}} \nonumber\\
&&\qquad\;\,~~~~~~={\Gamma_{{\rm{SM}}}^{h_{2}}\over \Gamma_{{\rm{NP}}}^{h_{2} }}\:{\Gamma_{{\rm{NP}}}(h_{2} \rightarrow gg) \over \Gamma_{{\rm{SM}}}(h_{2} \rightarrow gg)} \: |g_{{h_{2} VV}}|^2,\nonumber\\
&&\mu_{f\bar{f}}^{{\rm{VBF}}}(125) \approx {\Gamma_{{\rm{NP}}}(h_{2} \rightarrow{VV^*}) \over
\Gamma_{{\rm{SM}}}(h_{2} \rightarrow{VV^*})} \:{\Gamma_{{\rm{NP}}}(h_{2} \rightarrow f\bar{f}) /\Gamma_{{\rm{NP}}}^{h_{2}} \over \Gamma_{{\rm{SM}}}(h_{2} \rightarrow f\bar{f}) /\Gamma_{{\rm{SM}}}^{h_{2}}} \nonumber\\
&&\qquad\;\,~~~~~~={\Gamma_{{\rm{SM}}}^{h_{2}}\over \Gamma_{{\rm{NP}}}^{h_{2}}}\: |g_{{h_{2} VV}}|^2 \: |g_{{h_{2}ff}}|^2\qquad (V=Z,W;~f=b,\tau),
\label{signals}
\end{eqnarray}
with  ${\Gamma_{{\rm{NP}}}(h_{2} \rightarrow{VV^*}) \over
\Gamma_{{\rm{SM}}}(h_{2} \rightarrow{VV^*})} = |g_{{h_{2} VV}}|^2$ and ${\Gamma_{{\rm{NP}}}(h_{2} \rightarrow f\bar{f}) \over \Gamma_{{\rm{SM}}}(h_{2} \rightarrow f\bar{f}) } =|g_{{h_{2} ff}}|^2$.

\section{Excess at 95 GeV}
The signal strengths of the 95 GeV scalar excess in the $U(1)_X$VLFM are defined as the ratios of the corresponding production cross sections and branching ratios to their SM expectations,
\begin{eqnarray}
&&\mu_{\gamma\gamma}(95)={ \sigma_{{\rm{NP}}}({gg\rightarrow h_{1}})\over
\sigma_{{\rm{SM}}}({gg\rightarrow h_{1}})}\:{{\rm{BR}}_{{\rm{NP}}}(h_{1} \rightarrow\gamma\gamma)\over
{\rm{BR}}_{{\rm{SM}}}(h_{1} \rightarrow\gamma\gamma)} \nonumber\\
&&~~~~~~~~~~\approx{\Gamma_{{\rm{SM}}}^{h_{1}}\over \Gamma_{{\rm{NP}}}^{h_{1} }}\:{\Gamma_{{\rm{NP}}}(h_{1} \rightarrow gg) \over
\Gamma_{{\rm{SM}}}(h_{1} \rightarrow gg)}\:{\Gamma_{{\rm{NP}}}(h_{1} \rightarrow\gamma\gamma)\over
\Gamma_{{\rm{SM}}}(h_{1} \rightarrow\gamma\gamma)}, \nonumber\\
&&\mu_{b\bar{b}}(95)={ \sigma_{{\rm{NP}}}({Z^*\rightarrow Zh_{1}})\over
\sigma_{{\rm{SM}}}({Z^*\rightarrow Zh_{1}})}\:{{\rm{BR}}_{{\rm{NP}}}(h_{1} \rightarrow{b\bar{b}})\over
{\rm{BR}}_{{\rm{SM}}}(h_{1} \rightarrow{b\bar{b}})}\nonumber\\
&&~~~~~~~~~~\approx{\Gamma_{{\rm{SM}}}^{h_{1}}\over \Gamma_{{\rm{NP}}}^{h_{1} }}\:{\Gamma_{{\rm{NP}}}(h_{1}\rightarrow b\bar b) \over \Gamma_{{\rm{SM}}}(h_{1} \rightarrow b\bar b)} \: |g_{{h_{1} ZZ}}|^2,
\label{mu}
\end{eqnarray}
with
\begin{eqnarray}
&&\Gamma_{{\rm{NP}}}^{h_{1}}\approx\sum\limits_{f=b,\tau,c} \Gamma_{{\rm{NP}}}(h_{1} \rightarrow f\bar{f})+\: \Gamma_{{\rm{NP}}}(h_{1}\rightarrow gg),
\end{eqnarray}

In the narrow width approximation, the signal strengths normalized to the SM predictions are given by Eq. \ref{mu}. "NP" refers to the predictions in our BSM model ($U(1)_X$VLFM), and "SM" refers to the hypothetical SM Higgs with the same mass~\cite{n1}.
$\Gamma_{{\rm{NP}}}^{h_{1}}$ denotes the total decay width of the 95 GeV scalar in the $U(1)_X$VLFM, while $\Gamma_{{\rm{SM}}}^{h_{1}}$ corresponds to the total width of a SM Higgs boson with the same mass. The partial decay widths $\Gamma_{{{\rm{NP}}}({{\rm{SM}}})}(h_{1} \rightarrow gg, \gamma\gamma, f\bar{f})$ are evaluated analogously to Eqs.~(\ref{hgg}), (\ref{hrr}) and (\ref{hff}), with $h_{2} $ replaced by $h_{1} $.

\section{numerical analysis}
In our numerical analysis, we assume that the lightest CP-even Higgs boson corresponds to a state with a mass around $95~\rm GeV$, while the next-to-lightest CP-even Higgs boson is identified with the $125~\rm GeV$ SM-like state. The masses and associated signal strengths of these states are then discussed within the $U(1)_X$VLFM framework, subject to the following phenomenological constraints:

1. The second-lightest CP-even Higgs boson $h_{2} $ is interpreted as the observed Higgs boson, with a mass of $m_{h_{2}} = 125.20 \pm 0.11~{\rm GeV}$ as reported by the PDG~\cite{PDG}.

2. The third-generation fermion masses are required to match the SM values after mixing with the vector-like fermions~\cite{PDG}.

3. For the vector-like fermions in the $U(1)_X$VLFM, we take them at TeV order~\cite{Benbrik:2024fku,CMS:2025urb,ATLAS:2025wgc}.

4. Motivated by previous analyses of $U(1)_X$ extensions with vector-like fermions, we allow a TeV-scale $Z'$ mass in our numerical study, as such values are shown to be phenomenologically viable in Refs.~\cite{Xu:2018pnq,Dinh:2023ezl}.

All these constraints are applied in the parameter scan to ensure consistency with current experimental data. For the subsequent numerical analysis, we select the following representative benchmark scenario:
\begin{eqnarray}
&&Q_a = 1,~~Q_b = 1,~~ Y_{XN} = 0.4,~~ Y_{PN} = 0.4,~~ \lambda_H = -0.12,~~ \lambda_P =-0.003,~~\nonumber\\&&\lambda_X = -0.05,~~\lambda_{HP} = -0.01,~~\lambda_{HX} = -0.03,~~\lambda_{PX} = -0.01,~~ \nonumber\\&&
Y_{u_3} = 1.51Y_t,~~Y_{d_3}=1.54Y_b,~~Y_{e_3} = 1.5Y_\tau.
\end{eqnarray}
Here $Y_t$, $Y_b$, and $Y_\tau$ denote the Yukawa coupling of the top quark, bottom quark, and tau lepton, respectively, defined as
$Y_t = \sqrt{2} m_t/v$, $Y_b = \sqrt{2} m_b/v$, and $Y_\tau = \sqrt{2} m_\tau/v$.

The benchmark points in this work are not chosen arbitrarily, but are determined by the model structure and experimental constraints. In the $U(1)_X$VLFM model considered in this paper, the CP-even neutral scalar sector is described by a $3\times3$ mass matrix, and it is required to simultaneously realize a $125.20 \pm 0.11~{\rm GeV}$ scalar consistent with the SM Higgs and a light scalar around 95 GeV~\cite{PDG}. These two mass conditions impose very strong constraints on the parameter space, such that the parameters directly related to the scalar masses are strongly restricted to relatively narrow intervals, thereby significantly reducing their freedom.

On this basis, we further combine the Higgs signal strength measurements at 125 GeV and the excess around 95 GeV, and perform a joint $\chi^2$ fit over multiple decay channels. This procedure imposes additional experimental constraints on the parameter space, such that the region satisfying all observations is further reduced.

In contrast, those parameters that are not directly related to the masses of these two scalars, but mainly affect their decay properties, have relatively wider allowed ranges and retain some degree of variation. Although the benchmark point we present corresponds to the optimal fit obtained from the $\chi^2$ analysis, these parameters can still vary within a certain range while maintaining a good description of the experimental data.

The benchmark point given in this work corresponds to the best-fit point obtained from the $\chi^2$ analysis. Therefore, the apparent concentration of these parameter values mainly originates from the compression of the parameter space required to simultaneously satisfy the 95 GeV and 125 GeV experimental constraints, rather than from any artificial choice. In addition, the constrained parameter ranges are also consistent with the typical physical scales expected in the model.

Using this benchmark, the $U(1)_X$VLFM predictions are tested against Higgs data via a $\chi^2$ analysis, where points with $\chi^2$ within the $3\sigma$ range of the best-fit value are considered favored. The $\chi^2$ is defined as
\begin{eqnarray}
&&\chi^2 =\sum_i \left(\frac{\mu^{th}_i - \mu^{exp}_i}{\delta_i}\right)^2\nonumber\\&&
~~~= \Big(\frac{m^{th}_{h(125)} - m^{exp}_{h(125)}}{\delta_{m_{h}(125)}}\Big)^2 + \Big(\frac{\mu^{th}_{b\bar b(95)} - \mu^{exp}_{b\bar b(95)}}{\delta_{\mu^{95}_{b\bar b}}}\Big)^2
+ \Big(\frac{\mu^{th}_{\gamma\gamma(95)} - \mu^{exp}_{\gamma\gamma(95)}}{\delta_{\mu^{95}_{\gamma \gamma}}}\Big)^2 \nonumber\\&&
~~ ~+ \Big(\frac{\mu^{th}_{\gamma\gamma(125)} - \mu^{exp}_{\gamma\gamma(125)}}{\delta_{\mu^{125}_{\gamma \gamma}}}\Big)^2
+ \Big(\frac{\mu^{th}_{b\bar b(125)} - \mu^{exp}_{b\bar b(125)}}{\delta_{\mu^{125}_{b\bar b}}}\Big)^2
+ \Big(\frac{\mu^{th}_{ZZ^*(125)} - \mu^{exp}_{ZZ^*(125)}}{\delta_{\mu^{125}_{ZZ^*}}}\Big)^2\nonumber\\&&
~~ ~ + \Big(\frac{\mu^{th}_{WW^*(125)} - \mu^{exp}_{WW^*(125)}}{\delta_{\mu^{125}_{WW^*}}}\Big)^2
+ \Big(\frac{\mu^{th}_{\tau\bar\tau(125)} - \mu^{exp}_{\tau\bar\tau(125)}}{\delta_{\mu^{125}_{\tau\bar\tau}}}\Big)^2,\label{kf}
\end{eqnarray}
Here,  $\mu^{exp}_i$ denotes the experimental value of the corresponding observable, while $\mu^{th}_i$ represents its theoretical prediction. The term $\delta_i$ stands for the total uncertainty of the observable, which includes statistical and systematic experimental errors as well as theoretical uncertainties. The fit includes the $125~\text{GeV}$ Higgs boson mass, its signal strengths in the $\gamma\gamma$, $ZZ^*$, $WW^*$, $b\bar b$, and $\tau\bar\tau$ channels, as well as the potential excess at $95~\text{GeV}$ in the $\gamma\gamma$ and $b\bar b$ final states. By taking into account multiple experimental constraints from various physical processes, the parameter space obtained through the $\chi^2$ analysis becomes more reasonable and accurate.

\begin{figure}[ht]
\setlength{\unitlength}{5mm}
\centering
\includegraphics[width=2.5in]{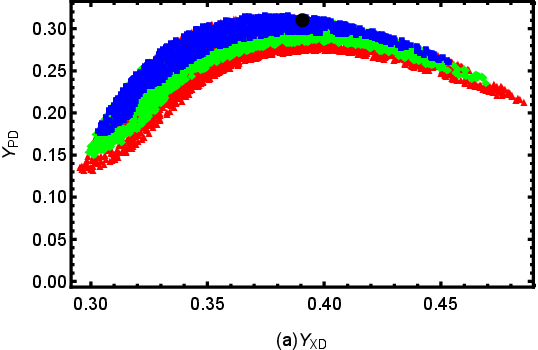}
\setlength{\unitlength}{5mm}
\centering
\includegraphics[width=2.4in]{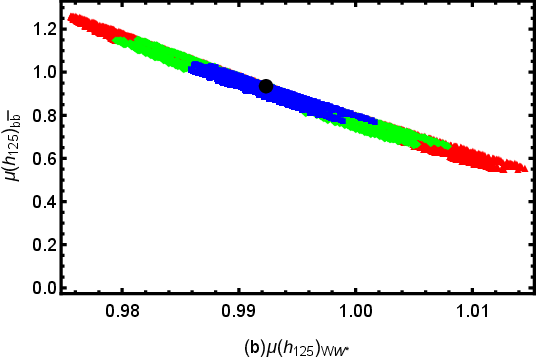}
\setlength{\unitlength}{5mm}
\centering
\includegraphics[width=2.5in]{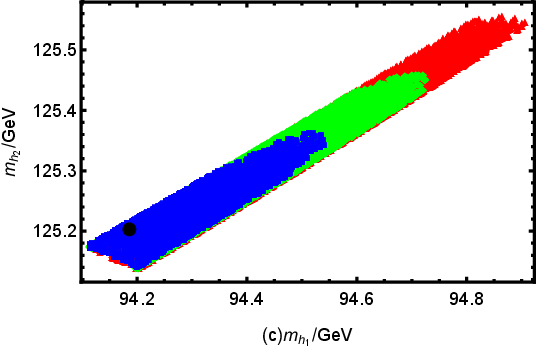}
\caption{The symbols indicate the $\chi^2$ confidence regions: $\bullet$ marks the best-fit point ($\chi^2_{\rm min} = 3.83$), $\textcolor{blue}{\blacksquare}$ the $1\sigma$ region ($\chi^2 \leq 6.13$), $\textcolor{green}{\blacklozenge}$ the $1$-$2\sigma$ region ($6.13 < \chi^2 \leq 10.01$), and $\textcolor{red}{\blacktriangle}$ the $2$-$3\sigma$ region ($10.01 < \chi^2 \leq 15.66$).} {\label {tu1}}
\end{figure}

The parameters $Y_{XD}$ and $Y_{PD}$ represent the Yukawa couplings that link the SM-like and vector-like down-type quarks. Specifically, $Y_{XD}$ connects the SM right-handed down quarks ($d_R$) with the vector-like left-handed components ($d_{XL}$), while $Y_{PD}$ couples the left- and right-handed vector-like quarks ($d_{XL}$ and $d_{XR}$). After the singlet scalars $S$ and $\phi_P$ acquire VEVs $v_S$ and $v_P$, these couplings contribute to the mass mixing structure of the extended quark sector, affecting the quark masses as well as their interactions with the Higgs and $Z'$ bosons. They are therefore taken as scan parameters in our analysis. Fig.~\ref{tu1} shows the results of the parameter scan with the benchmark set $g_X=0.41$, $g_{YX}=-0.1$, $Y_{PU}=0.41$, $Y_{XU}=0.49$, $Y_{PE}=0.32$, $Y_{XE}=0.46$, $v_S=1600~\rm GeV$, and $v_P=1900~\rm GeV$. The scan ranges are $0<Y_{XD}<0.8$ and $0.1<Y_{PD}<0.6$. The best-fit point ($\chi^2_{\rm min}=3.83$) is marked by a black dot, while the $1\sigma$, $2\sigma$, and $3\sigma$ confidence regions are represented by $\textcolor{blue}{\blacksquare}$, $\textcolor{green}{\blacklozenge}$, and $\textcolor{red}{\blacktriangle}$, respectively.

As shown in Fig.~\ref{tu1}(a), the allowed region in the $(Y_{XD},Y_{PD})$ plane forms a narrow arc-shaped band, indicating a strong correlation between the two parameters. The viable points cluster in the ranges $Y_{XD}\sim0.30\text{-}0.47$ and $Y_{PD}\sim0.15\text{-}0.30$, with the best-fit point near (0.391,0.310). The $1\sigma$ ($\textcolor{blue}{\blacksquare}$) region appears on the outer side of the arc and is somewhat shorter in arc length, while the $2\sigma$ ($\textcolor{green}{\blacklozenge}$) and $3\sigma$ ($\textcolor{red}{\blacktriangle}$) regions lie inward along the same trajectory and cover progressively larger portions. Within this correlated region, lower $\chi^2$ values are more often found at relatively higher $Y_{PD}$, suggesting a mild preference for larger $Y_{PD}$.

In Fig.~\ref{tu1}(b), the points form a narrow, descending band, revealing a clear anticorrelation between $\mu(h_{125})_{WW^*}$ and $\mu(h_{125})_{b\bar b}$. The lowest-$\chi^2$ points are shown in $\textcolor{blue}{\blacksquare}$, transitioning to $\textcolor{green}{\blacklozenge}$ and then $\textcolor{red}{\blacktriangle}$ toward the band edges. This pattern means that when one signal strength increases, the other must decrease accordingly. As a result, they cannot vary freely and are strongly constrained by the precision Higgs measurements.

In Fig.~\ref{tu1}(c), a diagonal band of allowed points is observed, indicating a clear positive correlation between the two Higgs masses. As $m_{h_1}$ increases from about 94.1 to 94.9~GeV, $m_{h_2}$ rises in parallel from roughly 125.1 to 125.6~GeV. The band is relatively broad, suggesting that both masses can shift together within a finite range while remaining consistent with the constraints. The best-fit point, indicated by the black dot, is located at approximately $m_{h_1}=94.19~\rm GeV$ and $m_{h_2}=125.20~\rm GeV$, a result particularly significant as it simultaneously aligns with the 95~GeV scalar excess and the observed 125~GeV Higgs boson.

\begin{figure}[ht]
\setlength{\unitlength}{5mm}
\centering
\includegraphics[width=2.5in]{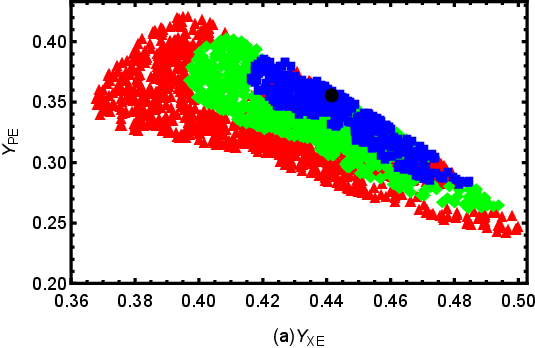}
\setlength{\unitlength}{5mm}
\centering
\includegraphics[width=2.5in]{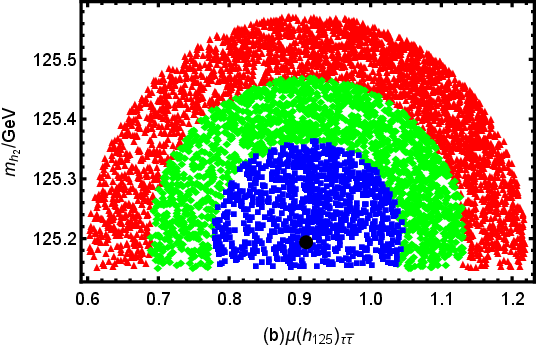}
\setlength{\unitlength}{5mm}
\centering
\includegraphics[width=2.5in]{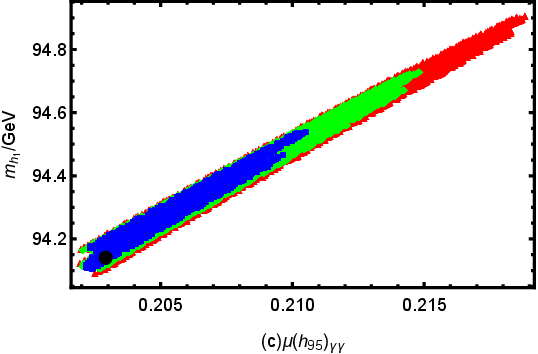}
\caption{The symbols indicate the $\chi^2$ confidence regions: $\bullet$ marks the best-fit point ($\chi^2_{\rm min} = 4.0$), $\textcolor{blue}{\blacksquare}$ the $1\sigma$ region ($\chi^2 \leq 6.3$), $\textcolor{green}{\blacklozenge}$ the $1$-$2\sigma$ region ($6.3 < \chi^2 \leq 10.18$), and $\textcolor{red}{\blacktriangle}$ the $2$-$3\sigma$ region ($10.18 < \chi^2 \leq 15.83$).} {\label {tu2}}
\end{figure}

The parameter $Y_{XE}$ governs the mixing between the vector-like lepton and the tau lepton, while $Y_{PE}$ primarily determines the mass of the vector-like lepton. Fig.~\ref{tu2} presents the $\chi^2$ distribution obtained from a scan over $Y_{XE}$ and $Y_{PE}$ within $0<Y_{XE}<0.8$ and $0.1<Y_{PE}<0.6$, using the benchmark parameters $g_X=0.41$, $g_{YX}=-0.1$, $Y_{PU}=0.41$, $Y_{XU}=0.49$, $Y_{PD}=0.3$, $Y_{XD}=0.41$, $v_S=1600\rm~GeV$, and $v_P=1900~\rm GeV$. The best-fit point ($\chi^2_{\rm min}=4.0$, $\bullet$) is indicated, with $1\sigma$ ($\textcolor{blue}{\blacksquare}$), $1$-$2\sigma$ ($\textcolor{green}{\blacklozenge}$), and $2$-$3\sigma$ ($\textcolor{red}{\blacktriangle}$) regions highlighting the parameter space favored by the fit.

In Fig.~\ref{tu2}(a), the allowed parameter space forms a tilted, wedge-like strip extending from the upper left to the lower right of the ($Y_{XE}$,$Y_{PE}$) plane, indicating a clear inverse correlation between the two couplings. The best-fit point lies within the compact 1$\sigma$ region, while the 2$\sigma$ and 3$\sigma$ regions expand outward along the same orientation. This stratified pattern suggests that only correlated changes of $Y_{XE}$ and $Y_{PE}$ are compatible with the Higgs signal strength measurements, thereby imposing stringent constraints on the Yukawa structure in the lepton sector.

In Fig.~\ref{tu2}(b), we map out the viable points in the plane defined by the $\tau\bar\tau$ signal strength of $h_{125}$ and the mass of the 125~GeV Higgs boson. The results exhibit a series of nested, semicircular layers centered around $m_{h_2}\simeq 125.2~\text{GeV}$, reflecting the strong experimental preference for a Higgs boson mass consistent with the SM-like Higgs. The $1\sigma$ region ($\textcolor{blue}{\blacksquare}$) is tightly concentrated near this central value, while the $2\sigma$ ($\textcolor{green}{\blacklozenge}$) and $3\sigma$ ($\textcolor{red}{\blacktriangle}$) regions form successive shells extending outward, indicating a gradual deterioration of the fit quality as the mass deviates from the preferred point. The best-fit point, marked by the black dot at $(\mu(h_{125})_{\tau\bar\tau}, m_{h_2}) \simeq (0.91, 125.19~\text{GeV})$, lies extremely close to the PDG world average of $m_{h_2}=125.20\pm0.11~\text{GeV}$, demonstrating excellent agreement between the model prediction and current Higgs measurements. This pattern illustrates that with the Higgs mass fixed close to its measured value, the $\tau\bar\tau$ signal strength is allowed to vary only within a very narrow range due to strong experimental constraints.

In Fig.~\ref{tu2}(c), the correlation between the diphoton signal strength of the 95~GeV scalar $\mu(h_{95})_{\gamma\gamma}$ and its mass $m_{h_1}$ is shown. The distribution of points forms an inclined band, indicating a positive correlation: $m_{h_1}$ increases from approximately 94.1~GeV to 95.0~GeV as $\mu(h_{95})_{\gamma\gamma}$ rises. The global best-fit point is located at the lower boundary of this region. The symbols and their colors represent the CL, with $\textcolor{blue}{\blacksquare}$ denoting the most favored parameter space, followed in sequence by $\textcolor{green}{\blacklozenge}$ and then $\textcolor{red}{\blacktriangle}$. This correlation reflects the intrinsic constraints within the model, which link the diphoton rate and mass of the 95~GeV scalar.

\begin{figure}[ht]
\setlength{\unitlength}{5mm}
\centering
\includegraphics[width=2.5in]{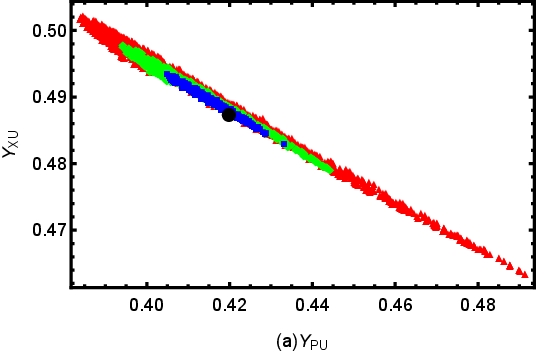}
\setlength{\unitlength}{5mm}
\centering
\includegraphics[width=2.5in]{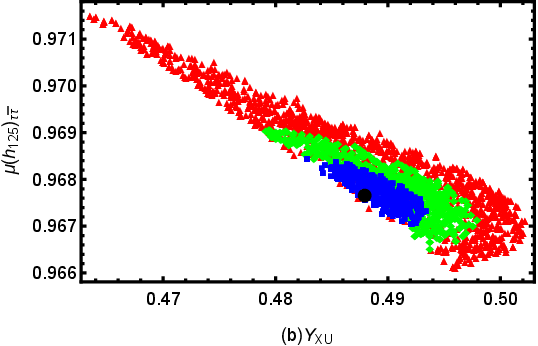}
\setlength{\unitlength}{5mm}
\centering
\includegraphics[width=2.5in]{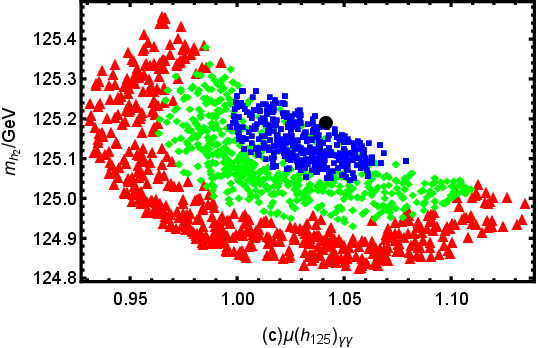}
\setlength{\unitlength}{5mm}
\centering
\includegraphics[width=2.5in]{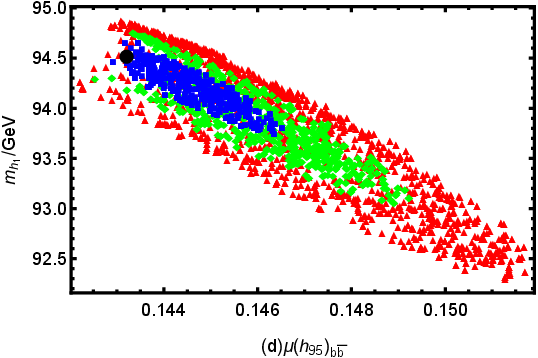}
\caption{The symbols indicate the $\chi^2$ confidence regions: $\bullet$ marks the best-fit point ($\chi^2_{\rm min} = 3.36$), $\textcolor{blue}{\blacksquare}$ the $1\sigma$ region ($\chi^2 \leq 5.66$), $\textcolor{green}{\blacklozenge}$ the $1$-$2\sigma$ region ($5.66 < \chi^2 \leq 9.54$), and $\textcolor{red}{\blacktriangle}$ the $2$-$3\sigma$ region ($9.54 < \chi^2 \leq 15.19$).} {\label {tu3}}
\end{figure}

Similarly, in the up-type quark sector, the parameters $Y_{XU}$ and $Y_{PU}$ characterize the Yukawa interactions responsible for the coupling between the SM top quark and its vector-like fermion. They determine the strength of this interaction, thereby influencing the formation of the up-type quark mass matrix and the mixing pattern in the extended quark sector. Both parameters are included as scan variables in the numerical analysis. In Fig.~\ref{tu3}, with $g_X=0.41$, $g_{YX}=-0.1$, $Y_{XE}=0.46$, $Y_{PE}=0.32$, $Y_{XD}=0.41$, $Y_{PD}=0.3$, $v_S=1600~\rm GeV$, and $v_P=1900~\rm GeV$ fixed, we perform a parameter scan over $Y_{XU}$ and $Y_{PU}$ within $0<Y_{XU}<0.8$ and $0.1<Y_{PU}<0.6$. The best-fit point is found at $\chi^2_{\rm min} = 3.36$, while the surrounding regions correspond approximately to $1\sigma$ ($\chi^2 \leq 5.66$), $1$-$2\sigma$ ($5.66 < \chi^2 \leq 9.54$), and $2$-$3\sigma$ ($9.54 < \chi^2 \leq 15.19$) CL.

Fig.~\ref{tu3}(a) displays the fitted parameter distribution in the $(Y_{PU},Y_{XU})$ plane. The points align along a well-defined descending strip, reflecting the compensating relation between the two couplings. The global best-fit point, marked by the black dot at $(Y_{PU},Y_{XU})\simeq(0.420,0.487)$, lies within the compact $1\sigma$ region, which is successively enclosed by the broader $2\sigma$ and $3\sigma$ confidence regions. This nested structure highlights the statistical hierarchy of the fit and demonstrates that Higgs observables can only be satisfied through a tightly correlated adjustment of $Y_{PU}$ and $Y_{XU}$.

Fig.~\ref{tu3}(b) shows the correlation between $Y_{XU}$ and $\mu(h_{125})_{\tau\bar\tau}$. Compared with Fig.~\ref{tu3}(a), the distribution here is noticeably wider, with the outer regions extending into a long descending tail. The best-fit point lies near the center of the innermost domain, and any deviation from this point quickly worsens the fit. This pattern shows that although some flexibility exists, the $\tau\bar\tau$ signal strength remains stringently correlated with $Y_{XU}$.

Fig.~\ref{tu3}(c) presents the distribution of points in the plane of the $h_{125}$ diphoton signal strength versus its mass ($\mu(h_{125})_{\gamma\gamma}$, $m_{h_2}$). The points trace a tilted, semicircular arc: $m_{h_2}$ is constrained within a narrow range of 124.8-125.4 GeV, while $\mu(h_{125})_{\gamma\gamma}$ varies close to the SM expectation, spanning 0.95-1.10. It reveals a compact $\textcolor{blue}{\blacksquare}$ core, successively surrounded by $\textcolor{green}{\blacklozenge}$ and $\textcolor{red}{\blacktriangle}$ layers that form nested semicircular envelopes, with the outer layer spanning the widest arc. The global best-fit point at ($\mu(h{_{125}})_{\gamma\gamma}, m_{h_2}) \simeq (1.04, 125.19~\mathrm{GeV})$ lies well inside the core. A mild anticorrelation emerges along the arc, with $m_{h_2}$ decreasing slightly as $\mu(h_{125})_{\gamma\gamma}$ increases, indicating constrained joint variations rather than independent scatter.

In Fig.~\ref{tu3}(d), the allowed points cluster into a diagonal band extending from the upper-left to the lower-right of the ($\mu(h_{95})_{b\bar b}$, $m_{h_1}$) plane. The strip exhibits a compact $\textcolor{blue}{\blacksquare}$ core is enclosed by $\textcolor{green}{\blacklozenge}$ and then $\textcolor{red}{\blacktriangle}$ regions, forming a slender corridor of viable points. A clear anticorrelation is observed: larger values of $\mu(h_{95})_{b\bar b}$ correspond to smaller $m_{h_1}$, indicating that the two observables are tightly constrained by the fit. The best-fit point, ($\mu(h_{95})_{b\bar b}, m_{h_1}) \simeq (0.143,94.51~\mathrm{GeV})$, lies near the upper boundary of the $\textcolor{blue}{\blacksquare}$ core, highlighting a preference for a slightly heavier $h_1$ mass accompanied by a reduced $b\bar b$ signal strength.

\begin{figure}[ht]
\setlength{\unitlength}{5mm}
\centering
\includegraphics[width=2.5in]{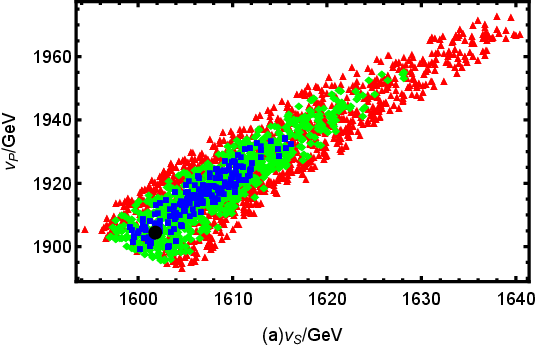}
\setlength{\unitlength}{5mm}
\centering
\includegraphics[width=2.4in]{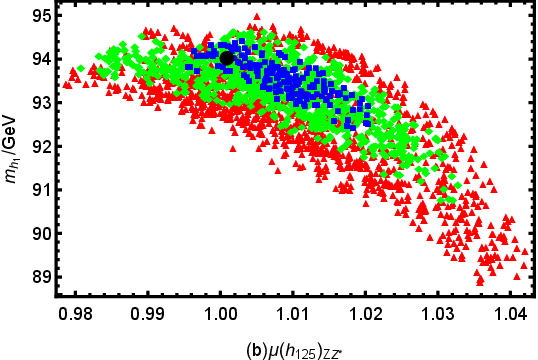}
\setlength{\unitlength}{5mm}
\centering
\includegraphics[width=2.5in]{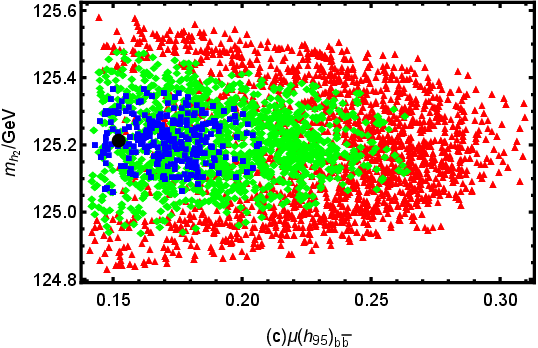}
\caption{The symbols indicate the $\chi^2$ confidence regions: $\bullet$ marks the best-fit point ($\chi^2_{\rm min} = 4.39$), $\textcolor{blue}{\blacksquare}$ the $1\sigma$ region ($\chi^2 \leq 6.69$), $\textcolor{green}{\blacklozenge}$ the $1$-$2\sigma$ region ($6.69 < \chi^2 \leq 10.57$), and $\textcolor{red}{\blacktriangle}$ the $2$-$3\sigma$ region ($10.57 < \chi^2 \leq 16.22$).} {\label {tu4}}
\end{figure}

In the $U(1)_X$VLFM, $v_S$ and $v_P$ denote the VEVs of the singlet scalars $S$ and $\phi_P$, respectively. These VEVs break the extra $U(1)_X$ gauge symmetry and generate masses for the new gauge boson $Z'$ as well as the vector-like fermions through their Yukawa interactions. They also play an important role in determining the scalar mass spectrum. Therefore, we take $v_S$ and $v_P$ as key parameters in our numerical analysis. In Fig.~\ref{tu4}, we perform an extended scan over the scalar VEVs, $1200~{\rm GeV} < v_S < 2500~{\rm GeV}$ and $1600~{\rm GeV} < v_P < 2800~{\rm GeV}$, while keeping $g_X=0.41$, $g_{YX}=-0.1$, $Y_{XE}=0.46$, $Y_{PE}=0.32$, $Y_{XD}=0.41$, $Y_{PD}=0.3$, $Y_{XU}=0.49$, and $Y_{PU}=0.41$ fixed. The resulting $\chi^2$ distribution is shown, with the best-fit point at $\chi^2_{\rm min}=4.39$ ($\bullet$) and the $1\sigma$, $1$-$2\sigma$, and $2$-$3\sigma$ confidence regions indicated by $\textcolor{blue}{\blacksquare}$, $\textcolor{green}{\blacklozenge}$, and $\textcolor{red}{\blacktriangle}$, respectively.

Fig.~\ref{tu4}(a) displays the viable region in the $(v_S, v_P)$ plane. The points align along a narrow diagonal band, revealing a strong positive correlation between the two VEVs. The distribution is coded by $\chi^2$ using both color and shape, with a compact $\textcolor{blue}{\blacksquare}$ core, surrounded by $\textcolor{green}{\blacklozenge}$, and $\textcolor{red}{\blacktriangle}$ outer layers, corresponding to increasing $\chi^2$ values. The best-fit point, $(v_S, v_P)\simeq(1601.8,1915.2~\mathrm{GeV})$, lies near the center of the innermost region, representing the most statistically favored vacuum configuration.

Fig.~\ref{tu4}(b) depicts the correlation between the Higgs signal strength in the $ZZ^*$ channel and the mass of the lightest scalar $m_{h_1}$. The points trace a curved band with a negative slope, where larger $\mu(h_{125})_{ZZ^*}$ values correspond to smaller $m_{h_1}$. The best-fit point, $(\mu(h_{125})_{ZZ^*},m_{h_1}) \simeq (1.00,94.02~\mathrm{GeV})$, lies near the densest region of points. The distribution of points reflects the fit quality. Points near the center match the experimental data better, while those farther away correspond to a poorer fit along the same trend. Even small deviations of $\mu(h_{125})_{ZZ^*}$ from unity cause noticeable shifts in $m_{h_1}$, indicating that the light scalar sector is highly sensitive to precision Higgs measurements.

Fig.~\ref{tu4}(c) shows the distribution of points in the $(\mu(h_{95})_{\gamma\gamma}, m_{h_2})$ plane. The values of $\mu(h_{95})_{\gamma\gamma}$ span the interval $0.14 \lesssim \mu(h_{95})_{\gamma\gamma} \lesssim 0.31$, whereas the SM-like Higgs mass remains tightly constrained within a narrow window around $125~\mathrm{GeV}$. The points exhibit a layered structure rather than a uniform spread: a compact $\textcolor{blue}{\blacksquare}$ core is surrounded by a $\textcolor{green}{\blacklozenge}$ band and further enclosed by a $\textcolor{red}{\blacktriangle}$ envelope, delineating progressively less favored regions of parameter space. The global best-fit point is located at ($\mu(h_{95})_{\gamma\gamma}, m_{h_2})\simeq(0.152,\,125.21~\mathrm{GeV}$), lying within the innermost region. This pattern highlights that the Higgs mass is tightly constrained, in contrast to the greater flexibility of the diphoton rate.

\begin{figure}[ht]
\setlength{\unitlength}{5mm}
\centering
\includegraphics[width=2.5in]{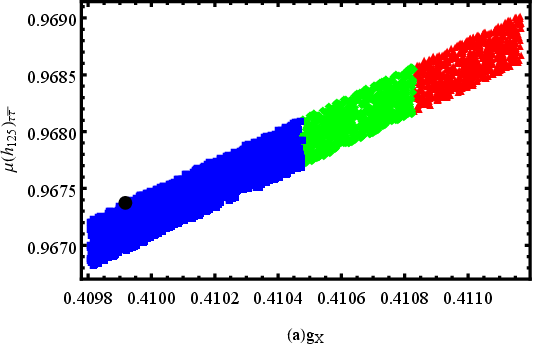}
\setlength{\unitlength}{5mm}
\centering
\includegraphics[width=2.5in]{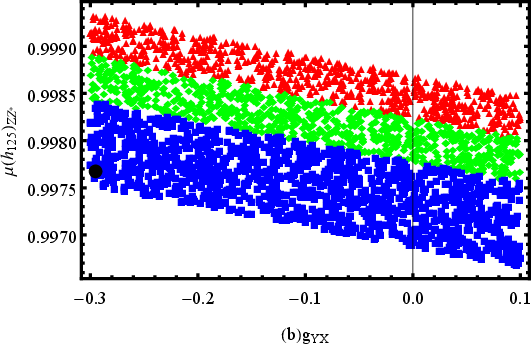}
\setlength{\unitlength}{5mm}
\centering
\includegraphics[width=2.5in]{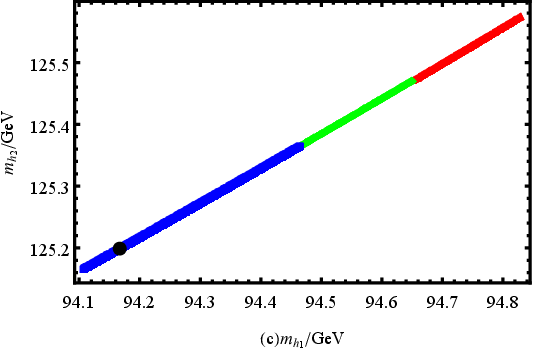}
\caption{The symbols indicate the $\chi^2$ confidence regions: $\bullet$ marks the best-fit point ($\chi^2_{\rm min} = 4.41$), $\textcolor{blue}{\blacksquare}$ the $1\sigma$ region ($\chi^2 \leq 6.71$), $\textcolor{green}{\blacklozenge}$ the $1$-$2\sigma$ region ($6.71 < \chi^2 \leq 10.59$), and $\textcolor{red}{\blacktriangle}$ the $2$-$3\sigma$ region ($10.59 < \chi^2 \leq 16.24$).} {\label {tu5}}
\end{figure}

In the $U(1)_X$VLFM, $g_X$ denotes the gauge coupling constant of the additional $U(1)_X$ symmetry. Parameter $g_{YX}$ denotes the gauge mixing between the hypercharge $U(1)_Y$ and the additional $U(1)_X$ symmetry. In Fig.~\ref{tu5}, instead of fixing $g_X$ and $g_{YX}$ as in the previous analyses, we allow them to vary within broader ranges, $0.25 < g_X < 0.7$ and $-0.3 < g_{YX} < 0.1$, keeping all other parameters constant: $Y_{XE}=0.46$, $Y_{PE}=0.32$, $Y_{XD}=0.41$, $Y_{PD}=0.3$, $Y_{XU}=0.49$, $Y_{PU}=0.41$, $v_S=1600~\rm GeV$, and $v_P=1900~\rm GeV$. The resulting distribution of points is displayed in terms of the $\chi^2$ values, with the best-fit point located at $\chi^2_{\rm min}=4.41$ (indicated by $\bullet$). The confidence regions are represented by $\textcolor{blue}{\blacksquare}$ ($1\sigma$, $\chi^2 \leq 6.71$), $\textcolor{green}{\blacklozenge}$ ($1$-$2\sigma$, $6.71 < \chi^2 \leq 10.59$), and $\textcolor{red}{\blacktriangle}$ ($2$-$3\sigma$, $10.59 < \chi^2 \leq 16.24$).

Fig.~\ref{tu5}(a) presents the correlation between $g_X$ and the Higgs signal strength $\mu(h_{125})_{\tau\tau}$. The scanned parameter points form a positively correlated band, indicating that larger values of $g_X$ generally lead to an enhancement of $\mu(h_{125})_{\tau\tau}$. The fit quality is represented through a combination of color and shape, where the most favored points ($\textcolor{blue}{\blacksquare}$) are concentrated in the lower-left region of the band, while less favored points ($\textcolor{green}{\blacklozenge}$ and $\textcolor{red}{\blacktriangle}$) are distributed further along the same correlation line.

Fig.~\ref{tu5}(b) plots $g_{YX}$ versus $\mu(h_{125})_{ZZ^*}$.
The points are distributed within a horizontally elongated band with a slight negative slope, with points within the $1\sigma$ region ($\textcolor{blue}{\blacksquare}$) located at the lower part of the band, followed by the $2\sigma$ ($\textcolor{green}{\blacklozenge}$) and $3\sigma$ ($\textcolor{red}{\blacktriangle}$) regions extending upward. This pattern highlights the sensitivity of $\mu(h_{125})_{ZZ^*}$ to gauge mixing effects.

Fig.~\ref{tu5}(c) shows the relationship between the masses of the two CP-even Higgs bosons, $m_{h_1}$ and $m_{h_2}$. The points are confined to a narrow diagonal strip, consistent with the scenario where one scalar has a mass near 95 GeV, aligning with the potential excess, and the other is compatible with the observed Higgs boson at 125 GeV. The fit quality is indicated by a color gradient along the diagonal, with the blue area encompassing the global best-fit point at ($m_{h_1}$,$m_{h_2}$)$\simeq$(94.17, 125.20 GeV), suboptimal points in green, and poorer fits in red. This blue-to-red color layering clearly illustrates the variation in $\chi^2$ and highlights the excellent agreement of the model predictions with the Higgs data.

In the parameter region corresponding to Fig.~4, $v_S = 1200\text{-}2500$ GeV and $v_P = 1600\text{-}2800$ GeV, while in Fig.~5, $g_{YX} = -0.3\text{-}0.1$ and $g_X = 0.25\text{-}0.7$. Since $\frac{g_{YX}^2}{g_X^2}<1$, the correction is dominated by the suppression factor $\frac{v^2}{16 v_P^2 + 4 v_S^2}$. For the representative parameter choices used in Figs.~4 and 5 ($g_{YX}=-0.1,~g_X=0.41,~v=246{~\rm GeV},~v_S=1600{~\rm GeV},~v_P=1900{~\rm GeV}$), the correction is at the $10^{-5}$ level.
\section{discussion and conclusion}
In this work, we perform a systematic study of the $U(1)_X$VLFM. The gauge symmetry of this framework is extended to $SU(3)_C \otimes SU(2)_L \otimes U(1)_Y \otimes U(1)_X$. Compared with the SM, the particle content is enlarged by three generations of right-handed neutrinos ($\nu_R$) and two singlet Higgs fields ($\phi$ and $S$). In addition, the model introduces one generation of vector-like quarks, vector-like lepton, and vector-like neutrino. With significantly fewer free parameters than supersymmetric extensions, the $U(1)_X$VLFM provides a theoretically consistent and economical framework for exploring NP beyond the SM.

In the scalar sector, the CP-even components of one Higgs doublet ($H$) and two Higgs singlets ($\phi$,~$S$) mix to form a $3\times 3$ mass-squared matrix. The lightest eigenstate can account for the 95 GeV excess reported at the LHC, whereas the second-lightest state requires one-loop corrections to reproduce the observed 125 GeV Higgs mass. Using the Higgs signal strengths measured in the $\gamma\gamma$, $ZZ^*$, $WW^*$, $b\bar b$, and $\tau\bar\tau$ channels by ATLAS and CMS, we perform a numerical fit of the model. Our results show that the predictions of the $U(1)_X$VLFM framework are in better agreement with the experimental data compared to the SM. Viable parameter points exist within the $1\sigma$ confidence interval of $\chi^2$ fitting, simultaneously accommodating both the 95 GeV excess and the 125 GeV Higgs boson. The gauge couplings $g_X$,~$g_{YX}$, singlet VEVs $v_S$,~$v_P$ and new Yukawa couplings $Y_{PD},Y_{XD},Y_{PE},Y_{XE},Y_{PU},Y_{XU}$ are found to be tightly constrained and highly sensitive to the numerical outcomes.

Furthermore, in the $U(1)_X$VLFM framework, the inert right-handed neutrino may satisfy the constraints from dark matter direct detection experiments, and the model may additionally provide a new avenue to address the observed lepton flavor universality violation at tree level. In our future work, we will further investigate the $U(1)_X$VLFM to determine its viable parameter space.

\begin{acknowledgments}
This work is supported by National Natural Science Foundation of China (NNSFC)
(No.12075074), Natural Science Foundation of Hebei Province
(A2023201040, A2022201022, A2022201017, A2023201041), Natural Science Foundation of Hebei Education Department (QN2022173), the Project of the China Scholarship Council (CSC) No. 202408130113. This work is also supported by Funda\c{c}\~{a}o para a Ci\^{e}ncia e a Tecnologia (FCT, Portugal) through the project UID/00777/2025 (https://doi.org/10.54499/UID/00777/2025).
\end{acknowledgments}

\appendix
\section{Analytic expressions}\label{A1}
Here, we take the down-type quark as an example and provide partial analytic expressions for illustration.
\begin{eqnarray}
&&\frac{\partial{V_d}}{\partial{\phi_H}} = -\frac{3}{16\pi^2}\Big[vY^2_{d_1}f(Q^2,m^2_{d_{1}})+vY^2_{d_2}f(Q^2,m^2_{d_{2}})+X_1f(Q^2,m^2_{d_{3}})+X_2f(Q^2,m^2_{d_{4}})\Big],
\nonumber\\&&
\frac{\partial{V_d}}{\partial{\phi_S}} = -\frac{3}{16\pi^2}\Big[X_3f(Q^2,m^2_{d_{3}})+X_4f(Q^2,m^2_{d_{4}})\Big], \nonumber\\&&
\frac{\partial{V_d}}{\partial{\phi_P}} = -\frac{3}{16\pi^2}\Big[X_5f(Q^2,m^2_{d_{3}})+X_6f(Q^2,m^2_{d_{4}})\Big],
\end{eqnarray}
with the auxiliary functions defined as
\begin{eqnarray}
&&f(Q^2,m^2_{d_i}) = 2m^2_{d_i}(\log{\frac{m^2_{d_i}}{Q^2}}-1),\nonumber\\&&
\Delta=\sqrt{(A-C)^2+4 B^2},~~Y_{d_1}=\frac{\sqrt{2}m_d}{v},~~Y_{d_2}=\frac{\sqrt{2}m_s}{v},\nonumber\\&&
A=\frac{1}{2}v^2 Y^2_{d_3},~~B=\frac{1}{2}v v_S Y_{d_3} Y_{XD},~~C=\frac{1}{2}v_S^2 Y_{XD}^2+\frac{1}{2}v_P^2 Y_{PD}^2,\nonumber\\&&
X_1=\frac{1}{2}v Y^2_{d_3}-\frac{1}{2\Delta}\Big[(A-C)v Y^2_{d_3}+2B v_S Y_{d_3} Y_{XD}\Big],\nonumber\\&&
X_2=\frac{1}{2}v Y^2_{d_3}+\frac{1}{2\Delta}\Big[(A-C)v Y^2_{d_3}+2B v_S Y_{d_3} Y_{XD}\Big],\nonumber\\&&
X_3=\frac{1}{2}v_S Y_{XD}^2-\frac{1}{2\Delta}\Big[(C-A)v_S Y_{XD}^2+2 B v Y_{d_3} Y_{XD}\Big],\nonumber\\&&
X_4=\frac{1}{2}v_S Y_{XD}^2+\frac{1}{2\Delta}\Big[(C-A)v_S Y_{XD}^2+2 B v Y_{d_3} Y_{XD}\Big],\nonumber\\&&
X_5=\frac{1}{2}v_P Y_{PD}^2-\frac{1}{2\Delta}\Big[(C-A)v_P Y_{PD}^2\Big],\nonumber\\&&
X_6=\frac{1}{2}v_P Y_{PD}^2+\frac{1}{2\Delta}\Big[(C-A)v_P Y_{PD}^2\Big].
\end{eqnarray}

\end{document}